\documentclass[12pt]{article}
\usepackage{amsmath}
\usepackage{algorithm}
\usepackage{algpseudocode}
\usepackage{graphicx,psfrag,epsf}
\usepackage{subcaption}
\usepackage{multirow}
\usepackage{enumerate}
\usepackage{enumitem}
\usepackage{microtype}
\usepackage{mathtools}
\usepackage[round]{natbib}
\usepackage{url} % not crucial - just used below for the URL 
\usepackage{amsfonts,mathrsfs,amsthm,amssymb}
\usepackage{comment}

\usepackage{array,booktabs,textcomp,stfloats}

\usepackage{setspace}
\usepackage[colorlinks=true, linkcolor=blue, citecolor=blue, urlcolor=blue]{hyperref}
\numberwithin{equation}{section}

\usepackage{xr-hyper}
\allowdisplaybreaks[3]

\newcommand{\blind}{1}

\newtheorem{thm}{Theorem}[section]

\newtheorem{corol}{Corollary}[section]
\newtheorem{prop}{Proposition}[section]
\newtheorem{example}{Example}[section]
\newtheorem{remark}{Remark}[section]
\newtheorem{assump}{Assumption}[section]
\newtheorem{definition}{Definition}[section]

\newcommand{\argmin}{\operatornamewithlimits{arg\,min}}

\DeclareMathOperator{\Ker}{Ker}
\DeclarePairedDelimiterX{\vvvert}[1]{\lvert\!\lvert\!\lvert}{\rvert\!\rvert\!\rvert}{#1}

\begin{document}

\def\spacingset#1{\renewcommand{\baselinestretch}%
{#1}\small\normalsize} \spacingset{1}

%%%%%%%%%%%%%%%%%%%%%%%%%%%%%%%%%%%%%%%%%%%%%%%%%%%%%%%%%%%%%%%%%%%%%%%%%%%%%%
\if1\blind
{
  \title{\bf A direct approach to tree-guided feature aggregation for high-dimensional regression}

  \author{%
    Jinwen Fu$^{1}$%,
    , Aaron J. Molstad$^{1,2}$\thanks{Corresponding author: \href{amolstad@umn.edu}{amolstad@umn.edu}}
    , and Hui Zou$^{1}$ \\
    {\normalsize
      $^{1}$School of Statistics, University of Minnesota, Minneapolis, MN 55455, USA}\\
      {\normalsize $^{2}$Department of Statistics, University of Florida, Gainesville, FL 32611, USA}
  }
  \date{}
  \maketitle
}
\fi
\if0\blind
{
  \bigskip
  \bigskip
  \bigskip
  \begin{center}
    {\LARGE\bf Tree-guided feature aggregation for high-dimensional regression and classification}
\end{center}
  \medskip
} \fi

\vspace{-10pt}
\begin{abstract}
In high-dimensional linear models, sparsity is often exploited to reduce variability and achieve parsimony. Equi-sparsity, where one assumes that predictors can be aggregated into groups sharing the same effects, is an alternative parsimonious structure that can be more suitable in certain applications. Previous work has clearly demonstrated the benefits of exploiting equi-sparsity in the presence of ``rare'' features \citep{yan2021rare}. In this work, we propose a new tree-guided regularization scheme for simultaneous estimation and feature aggregation. Unlike existing methods, our estimator avoids synthetic overparameterization and its detrimental effects. Novel techniques are developed to study the finite-sample error bound of this seminorm-induced regularizer under least squares and binomial deviance losses. Theoretically, compared to existing methods, the proposed method offers a faster or equivalent rate depending on the true equi-sparisty structure. Extensive simulation studies verify these findings. We show that our estimator can be computed very efficiently by exploiting special properties of our penalty. Finally, we illustrate the usefulness of the proposed method with an application to a microbiome dataset, where we conduct post-selection inference on the aggregated features' effects.
\end{abstract}

\noindent%

%\spacingset{1.45}
\spacingset{1.95} % DON'T change the spacing!
%\spacingset{1.25}
%\doublespacing
\vspace{-10pt}
\section{Introduction}
In microbiome research, modern high‑throughput sequencing experiments quantify the relative abundance of microbes in an environment (e.g., the human gut) by counting the number of rRNA reads sampled that can be assigned to a specific microbe or closely-related group of microbes (measured at the level of an operational taxonomic unit, OTU). For each subject, these experiments generate a vector $x\in\mathbb R^{p}$ with nonnegative entries and unit sum, where $p$ is the number of OTUs in the study. Naturally, the $j$th component of $x$ is the proportion of rRNA reads sampled attributable to the $j$th OTU, which is often interpreted as the relative abundance of that OTU in the subject's microbiome. 

A key analytical goal is to predict a subject-level trait $y$ (e.g., disease status) given $x$ using regression models whose coefficients are then interpreted in terms of associations between OTU abundance and the trait. The high-dimensionality and sparsity of $x$ (typically, many OTUs have zero rRNA reads) motivate researchers to impose sparsity on the coefficients, either by removing features before analysis or by using estimators that shrink coefficients toward zero. Such purely data‑driven feature selection, however, overlooks the biological structure of the OTUs and often complicates interpretation. Specifically, OTUs are embedded in a known taxonomic hierarchy reflecting phylogenetic similarity. A natural alternative is thus to aggregate OTUs by taxonomic rank and treat the resulting sums as predictors, representing the relative abundances of higher‑level taxa. The taxonomic tree (e.g., see Figure \ref{fig.exp6}) supports aggregation at various levels (e.g., genus, family, order), reducing dimensionality while yielding biologically meaningful groups, yet the aggregation level is usually chosen ad hoc. This paper introduces a tree‑guided method that adaptively selects aggregation levels based on the data and simultaneously estimates group effects relevant to a given trait.

%\textcolor{red}{Paragraph 2: The stucture in the solution is called equi-sparsity, and happens to have great prediction properties as an alternative variability reduction method. Give the two examples: RARE feature recovery result, and our latter proposition.}\\

%The task of feature aggregation is essentially forcing a group of predictors to share a common coefficient value, which is often referred to as \textit{equi-sparsity}\citep{she2010sparse}. The pursuit is philosophically similar to variable selection: one hopes to obtain an estimator with reduced variability and possibly more interpretable structure. But in the context of linear models, sparsity and equi-sparsity are different cures for collinearity, one tends to select a single feature from a group of highly correlated predictors(with a less understood mechanism), while the other views the group as a unity. In many cases, the latter might bring more benefits in both interpretability and theoretical properties. We give below two examples to illustrate the merits.\\
Feature aggregation compels a group of predictors to share a common coefficient, a property often called ``equi‑sparsity'' \citep{she2010sparse}. For example, in the linear model $y=x_1\alpha+x_2\beta+x_3\gamma+\epsilon$, assuming $\beta=\gamma$ induces the model $y=x_1\alpha+x_2\beta+x_3\beta+\epsilon=x_1\alpha+(x_2+x_3)\beta+\epsilon$ and aggregates the predictors $x_2$ and $x_3$. Like variable selection, feature aggregation can reduce variability and provide a more parsimonious model. However, in linear models, sparsity and equi‑sparsity address collinearity problem in distinct ways. A sparse procedure tends to pick a single representative from a group of highly correlated predictors---through a mechanism that is not always transparent---whereas equi‑sparsity treats the group as an indivisible unit. In many applications, the latter yields benefits both practically---through improved interpretability---and theoretically, as we discuss later.
 
Imposing an equi-sparse structure can sometimes achieve what sparsity cannot. \cite{yan2021rare} study a low-dimensional linear regression model in the presence of ``rare'' features. Here, the adjective rare refers to the feature being highly scarce in the training data, leading to many zero entries in the design matrix. In this setting, \cite{yan2021rare} show that the $L_1$-penalized regression estimator (henceforth, LASSO) fails to provide high-probability signed support recovery on the coefficients. As a remedy, they propose to aggregate groups of features sharing the same coefficient value, and show that LASSO then succeeds in support recovery on the aggregated features. 

In high-dimensional settings, the imposition of equi-sparsity serves as a dimension reduction tool that can provide improved prediction accuracy. In Section \ref{sect.oracla} of this paper, we compare the finite sample prediction error bounds of the ordinary least squares solution applied to the original features, and the ordinary least squares estimator applied to aggregated features. The result reveals a mechanism analogous to the sparse estimator, in that the decrement in prediction error comes from the reduction in the rank of the effective design matrix. We show that the advantage might persist even if the equi-sparsity structure only approximately holds. To be clear, we view sparsity and equi-sparsity as two roads that can lead to the goal of reduced variability and parsimony. Context should dictate which is more appropriate for a particular analysis. 

%It is only the scientist's choice to make in terms of which gives a better interpretation.

%\textcolor{red}{Paragraph 3: Talk about existing solutions, in two categories: (1) reparameterization; (2) direct regularization methods(fusion penalty; and another subtype that tends to select groups together). Keep the review brief with an emphasis on their drawbacks. }

In practice, when the true equi-sparsity structure is unknown, we must estimate which features to aggregate from a candidate structure given a priori. In our motivating microbiome data analysis, and in many other applications where equi-sparsity is appropriate, this structure is a tree. The existing literature on feature aggregation has taken two main approaches: pairwise fusion, or overparameterize-regularize. The fusion method penalizes the pairwise differences between coefficients to encourage precise equality. The vast literature is devoted to studying the weighting choices \citep{tibshirani2005sparsity,she2010sparse}, including some data-adaptive \citep{ke2015homogeneity,tang2016fused,chu2021adaptive} and graph-guided approaches \citep{chen2010graph}. Tree-guided weights, however, are notably absent from the literature, possibly because there is not a trivial way in which weights could ensure feature aggregation coheres with a given tree structure. Moreover, a challenge in fusion-based estimation is the intensive computation required to deal with the pairwise penalties, especially in high dimensions. 

Overparameterize-regularize methods typically operate by reparameterizing each regression coefficient as the sum of latent coefficients determined by a known structure. By imposing sparsity on the latent coefficients, the original coefficients can be naturally grouped to have the same values. \cite{yu2016sparse} reparameterize under the guidance of an undirected graph; while \cite{yan2021rare} and \cite{li2023s} designs the latent coefficients based on a tree, lying closer to our problem setting. These methods scale well to high-dimensional data, both computationally and theoretically. However, overparameterization arises several problems, including unfair selection and collinearity, as we will discuss in detail in both Section \ref{sect.method.conn.reparameterize} and \ref{sect.realdata}. Moreover, aggregation-driven penalties are naturally
seminorm-induced, as the shrinkage center is often a subspace rather than the origin. However,
existing works tend to side-step this complication, and only build theory for norm-induced penalties. As a result, no theoretical justification adequately fits the practical usage, especially in cases that adjust for other unpenalized covariates. 

In order to address the challenges, we propose a new, direct regularization scheme designed according to the given tree. The proposed method generalizes the aforementioned pairwise fusion penalty to allow larger groups. It avoids the problematic overparameterization of \cite{yan2021rare}, and only selects groups that strictly cohere with the tree structure. The benefits are verified in both theoretical and empirical results. We will introduce our method in Section \ref{sect.method}, and contrast it against existing methods. In Section \ref{sect.computation}, we discuss how to compute our estimator efficiently. In Section \ref{sect.theory}, we present finite-sample consistency results of the proposed estimator under high-dimensional regression and classification settings. We show that the unregularized subspace only impacts prediction error through its dimension in an $O_{P}(n^{-1})$ term. We show that our proposed method can have a better prediction error rate than the method of \citet{yan2021rare} (henceforth, RARE) under certain setting. Moreover, our theory provides insights into the influence of different tree structures on the prediction task and contributes to the literature by providing a novel treatment of seminorm-induced penalties. Simulation studies in Section \ref{sect.simulation} verify our theoretical findings and show a general advantage over RARE in various settings. Finally, in Section \ref{sect.realdata}, we apply our method to the analysis of microbiome data and illustrate its potential to produce new scientific insights. Before introducing our method, we use Section \ref{sect.oracla} to demonstrate the merit of feature aggregation.

\vspace{-10pt}
\section{Benefits of aggregation}\label{sect.oracla}
\vspace{-10pt}

To illustrate the benefit of feature aggregation, we first investigate the least-squares estimator under arbitrary aggregation in a high-dimensional regression setting. Let $y\in \mathbb R^n$ be the response and $X \in \mathbb{R}^{n\times p}$ be the design matrix. We assume the data-generating model 
$y=X\beta^*+\epsilon,$ where $\beta^* \in \mathbb{R}^p$ is the regression coefficient and $\epsilon$ is a  $\sigma$-sub-Gaussian vector with independent components. In this section, we show that the estimator using features aggregated according to the structure of $\beta^*$ can have prediction error bound smaller than that of the ordinary least-squares estimator. Moreover, the benefit of aggregation can persist even when the structure of $\beta^*$ only approximately holds.

%If we don't assume any structure on $\beta^*$ and use the ordinary least-squares estimator, $
   % \hat\beta_{\rm ols}\in\mathop{\rm argmin}_{\beta \in \mathbb{R}^p}\|y-X\beta\|_2^2,$
%then a finite-sample prediction error %bound is given by the following proposition.
\begin{prop}\label{prop.pl.ls}
    Assume $y = X\beta^* + \epsilon$ with $\epsilon$ having $n$ independent $\sigma$-sub-Gaussian entries. If $\mathop{\rm rank}(X)=r$, then with probability at least $1-2/r$,\vspace{-10pt}
    \begin{align}
        \frac{1}{\sqrt n}\|X\hat\beta_{\rm ols}-X\beta^*\|_2\le \frac{\sqrt{32\sigma^2\log(5) r+32\sigma^2\log (r)}}{\sqrt n}.
    \end{align}
\end{prop}
%A proof is provided in Appendix \ref{proof.prop}. 
We call the preceding result ``condition-free'' as it makes no assumptions on the design matrix $X$. The high-probability bound depends only on $\mathop{\rm rank}(X)=r$ and the noise variance $\sigma^2$. When $\beta^*$ is sparse, dropping zero-effect columns lowers $r$ and improves the bound. Likewise, an improvement can be made under equi-sparsity following a similar mechanism.

Now we assume an (approximately) correct equi-sparsity pattern is known. Let \(H=[h_1,\dots,h_K]\in\mathbb R^{p\times K}\) be the transformation that approximately maps the reduced
vector \(\tilde\beta^{*}\in\mathbb R^{K}\) to the full coefficient,
\(\beta^{*}\approx H\tilde\beta^{*}.\) Each column \(h_k\) is a binary vector encoding a set of features that are aggregated (i.e., share the same entries of $\beta^*$). 
For a given structure $H$, we define
\(
  \tilde\beta^{*}
  \in \arg\min_{\tilde\beta\in\mathbb R^{K}}
   \|X\beta^{*}-XH\tilde\beta\|_{2}^{2}, \) and the approximation error \(
  \gamma_{H}:=X\beta^{*}-XH\tilde\beta^{*}.
\) The error $\gamma_H$ vanishes when $\beta^{*}= H\tilde\beta^{*}$. With the grouped design \(\tilde X:=XH\), the model becomes $  y=\tilde X\tilde\beta^{*}+\gamma_{H}+\varepsilon .
$
We estimate \(\tilde\beta^{*}\) using aggregated least squares so that, ignoring the error $\gamma_H$, $
  \hat\beta_{\mathrm{als}}
  \in\arg\min_{\tilde\beta\in\mathbb R^{K}}\|y-\tilde X\tilde\beta\|_{2}^{2}.$
The next proposition reveals that the aggregated estimator has a smaller bound if the given structure is not far from the truth and the dimension reduction is ``sufficient''.
\begin{prop}\label{prop.agg.ls}
  Under the assumptions of Proposition \ref{prop.pl.ls},  we have $\mathop{\rm rank}(\tilde X)=\tilde r\le\min(r,K)$. Then, with probability at least $1-2/r$,\vspace{-10pt}
    \begin{align}
      \vspace{-10pt}
        \frac{1}{\sqrt{n}}\|XH\hat\beta_{\rm als}-X\beta^*\|_2\le \frac{\sqrt{32\sigma^2\log(5) \tilde r+32\sigma^2\log (r)}}{\sqrt n}+\frac{3\|\gamma_H\|_2}{\sqrt{n}}.
    \end{align}
\end{prop}
%The proof can be found in Appendix \ref{proof.prop.agg.ls}.
The result in Proposition \ref{prop.agg.ls} illustrates a trade-off between the variance controlled by aggregation and the bias caused by excessive aggregation. Specifically, the first term characterizes the variance, which is reduced compared to Proposition \ref{prop.pl.ls} if enough aggregation is conducted so that $\tilde r=K< r$. In the high-dimensional regime \(n\ll p\) (so \(r\le n\)),
``sufficient'' aggregation means choosing \(K<n\), forcing \(\tilde r=K<r\). The second term is the norm of the approximation error, representing the bias caused by excessive aggregation. Evidently, the aggregated estimator can be better even if equi-sparsity only holds approximately, especially when the noise variance $\sigma^2$ is large. This coheres with the findings of \cite{yan2021rare} in low-dimensional regression setting with normal noise.

\vspace{-10pt}
\section{Methodology}\label{sect.method}
\vspace{-10pt}
\subsection{Tree-guided convex regularization}
\vspace{-10pt}
When the true equi-sparsity structure is unknown, we assume that there is a tree $\mathcal T$ guiding the aggregation. With each leaf node denoting a feature, the tree gives information on the similarity between the feature effects. In the microbiome data analysis, for example, the taxonomic hierarchy naturally serves as $\mathcal T$. Biologically, it is reasonable to require that all features (OTUs) belonging to the same class be aggregated before those belonging to the same phylum, since each phylum contains many classes. Our goal is to discover an equi-sparsity structure that resides exactly on the tree hierarchy and is beneficial for the prediction task. To solve this problem, we first propose our tree-guided penalty, followed by a more detailed review of existing methods.

%The link function $f$ takes identity function $f(t)=t$ in the regression setting; and takes logit function $f(t)=1/(1+\exp(-t))$ in the classification setting. 

%\textcolor{red}{example of the projection: find the scalar $\beta$ such that $X_1\beta_1+X_2\beta_2+X_3\beta_3$ to be the closest to $(X_1+X_2+X_3)\beta$.}

%\subsection{Major Proposal: Tree-guided Penalty}

\begin{figure}[t]
\centering
\begin{subfigure}{0.32\textwidth}\centering
\includegraphics[width=\linewidth]{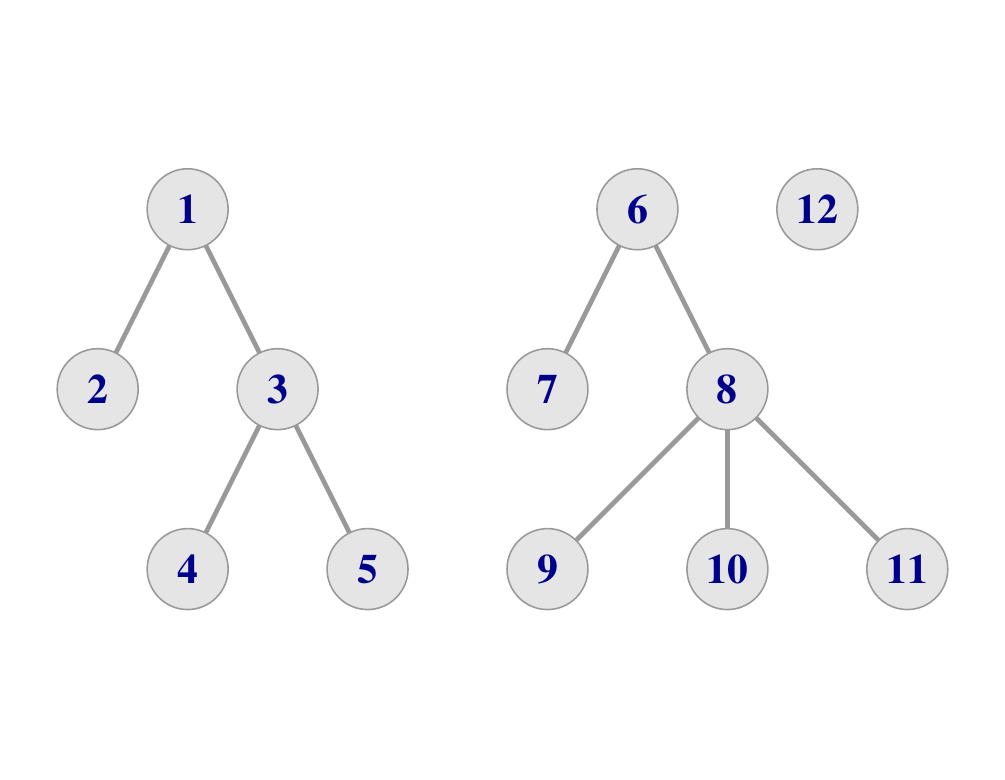}
\caption{}\label{fig.eg.forest}
\end{subfigure}
\begin{subfigure}{0.32\textwidth}\centering
\includegraphics[width=\linewidth]{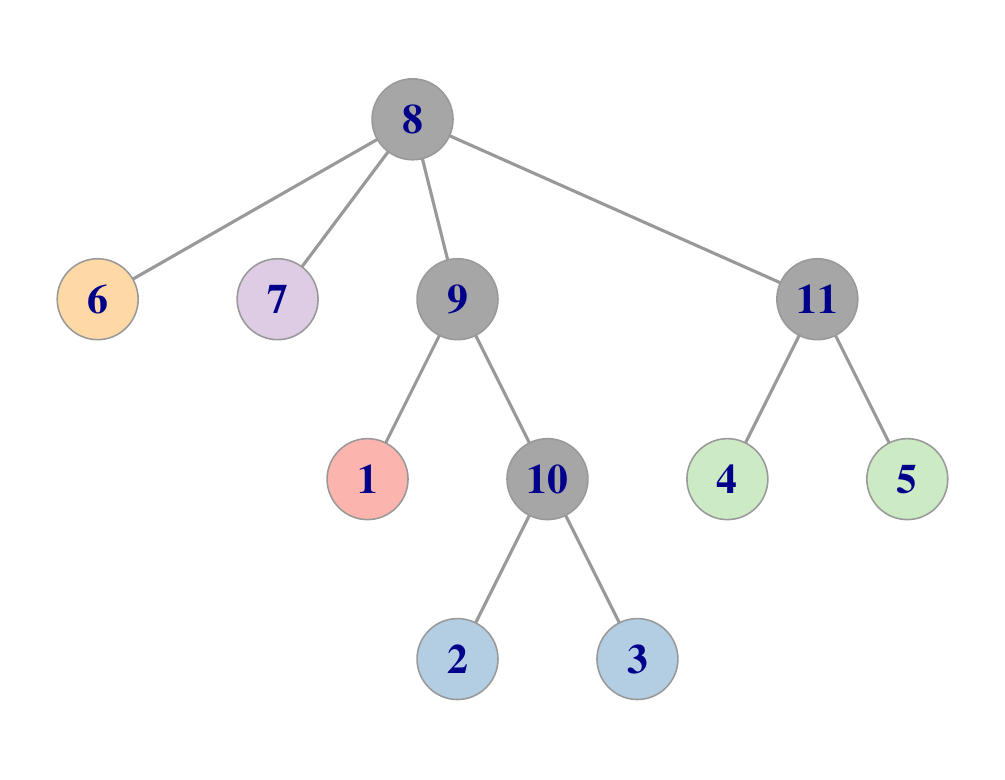}
\caption{}\label{fig.eg.tree}
\end{subfigure}\hfill
\begin{subfigure}{0.32\textwidth}\centering
\includegraphics[width=\linewidth]{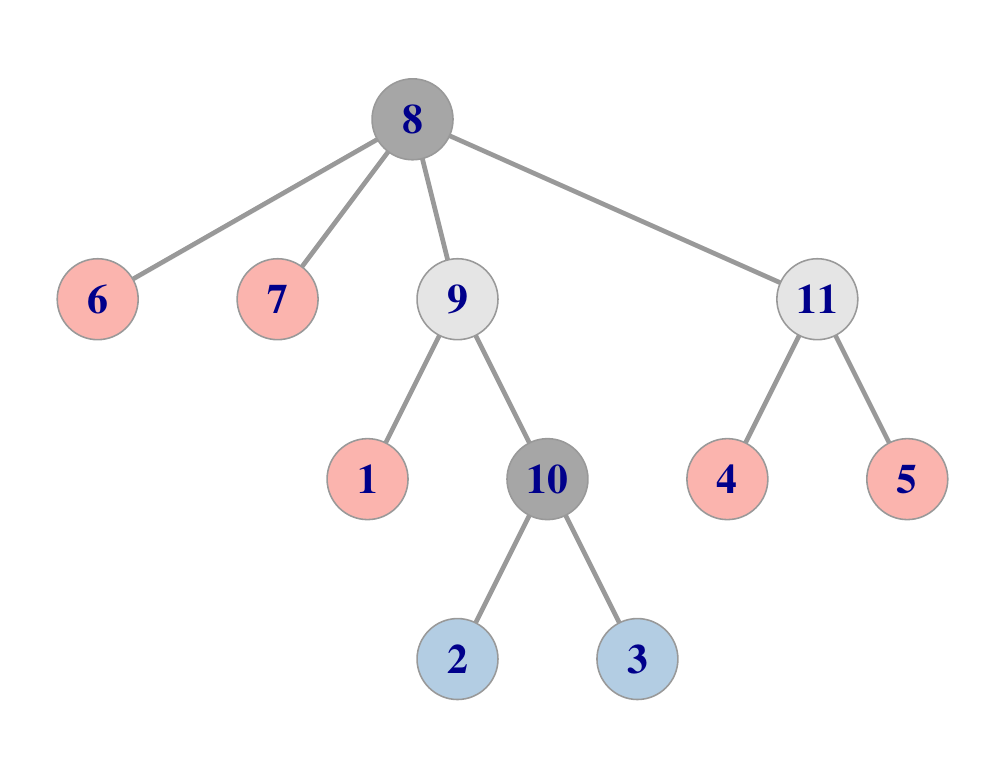}
\caption{}\label{fig.eg.wrong.tree}
\end{subfigure}
\caption{(a) An example of forest structure which can be handled by our method; (b) A tree-coherent partition where leaf nodes of the same color are aggregated; (c) A tree-incoherent partition where leaf nodes of the same color are aggregated.}\label{fig.eg}
\end{figure}

 Let $\{b_\ell\}_{\ell\in[L]}$ be all the nodes of a rooted tree (or forest) $\mathcal T$, and let $\{b_\ell\}_{\ell\in\mathcal I}$, for an index set $\mathcal I\subset[L]$, be the internal nodes (those with at least one child).  
The $p$ features correspond to the leaves $\mathcal L(\mathcal T)\subset\{b_\ell\}_{\ell\in[L]}$.  
For any node $b$, write $\mathcal T_b\subset\mathcal T$ for the subtree rooted at $b$ and
$\mathcal L(\mathcal T_b)\subset\mathcal L(\mathcal T)$ for its leaves, a candidate group of features.
We allow $\mathcal T$ to be a forest (see Figure \ref{fig.eg.forest}, for example) whose disconnected components must never be aggregated. Our goal is to discover an ``aggregating set'' that divides the leaf nodes into groups. We have the following definition from \citet[Definition 1,][]{yan2021rare}. 

\begin{definition}[\textbf{Aggregating set}]\label{def.agg.set}
A node set $\mathcal B=\{b_1,\dots,b_k\}\subset\mathcal T$ is an \emph{aggregating set} if
the leaf collections $\bigl\{\mathcal L(\mathcal T_{b_1}),\dots,\mathcal L(\mathcal T_{b_k})\bigr\}$ form a partition
$\mathcal A_{\mathcal B}=\{\mathcal A_1,\dots,\mathcal A_k\}$ of
$\mathcal L(\mathcal T)$, where $\mathcal A_\ell\subset[p]:=\{1,\dots,p\}$ indexes the $\ell$th group. If $b_\ell$ is itself a leaf, then $\mathcal L(\mathcal T_{b_\ell})=\{b_\ell\}$ is a singleton.
\end{definition}
Lemma 1 of \citet{yan2021rare} shows that every tree-coherent partition $\mathcal A$ admits a unique coarsest
aggregating set $\mathcal B$: finding a partition is then equivalent to choosing $\mathcal B$ and fixing a “resolution’’ of $\mathcal T$.
For example, Figure~\ref{fig.eg} gives two example partitions and represents groups in different colors. In Figure \ref{fig.eg.tree}, the tree-coherent partition $\mathcal A=\{\{6\},\{7\},\{1\},\{2,3\},\{4,5\}\}$ yields $\beta=(\beta_1,\beta_2,\beta_2,\beta_3,\beta_3,\beta_4,\beta_5)^\top$ and $\mathcal B=\{b_6,b_7,b_1,b_8,b_{10}\}$; whereas the partition $\mathcal A'=\{\{2,3\},\{1,4,5,6,7\}\}$ in Figure \ref{fig.eg.wrong.tree} is incompatible with the tree. We call an incompatible partition tree-incoherent.

To select the aggregating set in a data-driven way and simultaneously learn the effect vector $\beta$, we propose a tree-guided regularization method. If $g:\mathbb{R}^p \to \mathbb{R}$ is the loss function depending on $n$ data points, $\{(X_i, y_i)\}_{i=1}^n$,
we estimate the effect vector and aggregating set jointly by  \vspace{-10pt}
\begin{equation}\label{beta-hat}
    \vspace{-10pt}
\hat\beta\in \argmin_{\beta\in\mathbb R^p}\left\{g(\beta)+\lambda\,
\Omega_{\mathcal T,w}(\beta)\right\},
\end{equation}
with tuning parameter $\lambda>0$ and the hierarchical penalty  \vspace{-10pt}
\begin{equation}\label{penalty}
  \vspace{-10pt}
\Omega_{\mathcal T,w}(\beta)\;=\;\sum_{\ell\in\mathcal I}
w_\ell\min_{c\in\mathbb R}\bigl\|\beta_{\mathcal A_\ell}-c\,\mathbf 1_{a_\ell}\bigr\|_2
\;=\;\sum_{\ell\in\mathcal I}w_\ell\bigl\|D_{a_\ell}\beta_{\mathcal A_\ell}\bigr\|_2,
\end{equation}
where $w_\ell \geq 0$ is a user-specified weight, $\mathcal{A}_{\ell}$ indexes the leaf nodes corresponding to a subtree $\mathcal{T}_{b_\ell}$ rooted at the node $b_\ell$, $a_\ell=|\mathcal{A}_\ell|$ is the cardinality, $D_{a_\ell}=\mathbf{I}_{a_\ell} - \mathbf{1}_{a_\ell} \mathbf{1}_{a_\ell}^\top /a_l$ denotes a centering matrix with dimension $a_\ell$, and $\beta_{\mathcal{A}_\ell}$ is a subvector of $\beta$ indexed by $\mathcal{A}_\ell$. For each term, the non-differentiability at $D_{a_\ell}\beta_{\mathcal A_\ell}=0$ encourages
$\hat\beta_{\mathcal A_\ell}\propto\mathbf 1_{a_\ell}$, i.e. the exact aggregation of group $\mathcal A_\ell$.
The term
\(
\bigl\|D_{a_\ell}\beta_{\mathcal A_\ell}\bigr\|_2
=\bigl\|\beta_{\mathcal A_\ell}-\bar\beta_{\mathcal A_\ell}\mathbf 1_{a_\ell}\bigr\|_2
=\sqrt{a_\ell}\,{\rm se}\bigl(\beta_{\mathcal A_\ell}\bigr)
\)
can be interpreted as the (unscaled) group-wise estimated standard error. Thus, is it natural to set 
$w_\ell=a_\ell^{-1/2}$ to penalize the scaled variability within each group.

This penalty sums up the terms for all internal nodes $\{b_\ell\}_{\ell\in\mathcal I}$ in the tree. For example, in Figure \ref{fig.eg.tree}, there are only four internal nodes $\{b_{11},b_9,b_8,b_{10}\}$, corresponding to the groups \(\{\mathcal{A}_{{11}}=\{1,2,3,4,5,6,7\},\mathcal{A}_{9}=\{1,2,3\},\mathcal{A}_{{8}}=\{2,3\},\) \(\mathcal{A}_{{10}}=\{4,5\}\}\). Then the penalty is 
$
\Omega_{\mathcal{T},w}(\beta)= 7^{-1/2} \|D_{7} \beta_{\mathcal{A}_{{11}}} \|_2+ 3^{-1/2}\|D_{3} \beta_{\mathcal{A}_{9}} \|_2 +2^{-1/2}\|D_{2} \beta_{\mathcal{A}_{{8}}} \|_2 +2^{-1/2}\|D_{2} \beta_{\mathcal{A}_{{10}}} \|_2.$
The penalty is inspired by \cite{molstad2023multiresolution}, where they considered a categorical regression setting, and regularized the coefficients to enforce across-category homogeneity.

The benefit brought by the hierarchical penalty is immediately apparent: for any two groups, they are either nested or disjoint, so the direct selection of groups guarantees our aggregating structures are always tree-coherent. By solving \eqref{beta-hat}, we automatically obtain an aggregating set $\mathcal{B}$ and an aggregated estimator $\hat\beta$. We show that the degree of aggregation is determined by the penalty strength $\lambda$, which is selected by cross-validation in practice.

In Section \ref{sect.computation}, we develop an accelerated proximal gradient algorithm \citep{beck2009fast}. The proximal operator of
$\Omega_{\mathcal T,w}$ can be computed with a one-pass, non-iterative algorithm, so \eqref{beta-hat} can be solved efficiently. 

%\subsection{Connection with existing works}

%We first review the two categories closely related to our task and method in Section \ref{sect.method.conn.reparameterize} and \ref{sect.method.conn.regularize}. The last part \ref{sect.method.conn.other} also summarizes the methods that are loosely related to our task, but align philosophically with our pursuit.
\vspace{-10pt}
\subsection{Connection to overparameterize-regularize methods}\label{sect.method.conn.reparameterize}
%We first review the two categories closely related to our task and method in Section \ref{sect.method.conn.reparameterize} and \ref{sect.method.conn.regularize}. The last part \ref{sect.method.conn.other} also summarizes the methods that are loosely related to our task, but align philosophically with our pursuit.
% \subsubsection{reparameterize-regularize methods}\label{sect.method.conn.reparameterize}
\citet{yan2021rare} induce aggregation through a tree-based reparameterization.
They label the $i$th node in the tree with a latent coefficient $\gamma_i$ for each $i \in [L]$. For a leaf $j$, its coefficient $\beta_j$ is the sum of the latent coefficients of its ancestors, e.g. in Figure \ref{fig.eg.tree},
$
    \beta_1=\gamma_{8}+\gamma_9+\gamma_1,
    \beta_2=\gamma_{8}+\gamma_9+\gamma_{10}+\gamma_2, 
    \beta_3=\gamma_{8}+\gamma_9+\gamma_{10}+\gamma_3,$ and so on.
This can be summarized as $\beta=A\gamma$ where the expansion matrix $A\in\mathbb R^{p\times L}$ has ones in the columns corresponding to each leaf’s ancestors and zeros elsewhere. They estimate $\gamma$ via a LASSO on the reparameterized model  \vspace{-10pt}
\[
  \vspace{-10pt}
    \hat\gamma_{\rm rare} \in \mathop{\rm argmin}_{\gamma \in\mathbb R^L} \left\{ g(A\gamma)+\lambda \sum_{i \in[L]\//\text{roots}}^L w_i|\gamma_i|\right\},~~\hat\beta_{\rm rare}=A\hat\gamma_{\rm rare},
\]
where $g$ is the chosen loss (squared error in linear regression) and the $l_1$ penalty excludes the root. Because $X\beta=XA\gamma$, each internal node acts as a synthetic predictor equal to the sum of its descendants. Returning to Figure \ref{fig.eg.tree}, with direct selection on synthetic predictors, shrinking $\gamma_2$ and $\gamma_3$ to zero forces $\beta_2=\beta_3=\gamma_8+\gamma_9+\gamma_{10}$; additionally, zeroing $\gamma_1,\gamma_{10}$ collapses $\beta_1,\ \beta_2$ and $\beta_3$ to a common value $\gamma_8+\gamma_9$, achieving data-driven aggregation. The solution is both neat and can be computed efficiently. Empirically, \citet{bien2021tree} report that this RARE strategy outperforms the tree-cut search of \citet{guinot2018learning} in both simulations and real data applications. %They further applied RARE to the Microbiome data in \cite{bien2021tree}.

Although a natural solution, RARE has several limitations.\\
\textbf{(i) Tree-incoherent selection}.  Shrinking arbitrary latent
coefficients can aggregate leaves that do not share a subtree
(e.g.\ Fig.~\ref{fig.eg.wrong.tree} when only $\gamma_{8},\gamma_{10}$ remain nonzero),
undermining tree-based interpretation. This issue is observed in our microbiome data analysis (Section~\ref{sect.realdata}).  
\cite{li2023s} impose a descendant group penalty on the reparameterized parameters to enforce coherence, but at the cost of heavier computation.\\
\textbf{(ii) Theory–practice gap}.  In practice, it is always the case that some latent coefficients are not penalized, e.g., those corresponding to roots in the forest as Figure \ref{fig.eg.forest}. However, both \cite{yan2021rare} and \cite{li2023s} assume every node to be regularized in their theoretical analyses. As such, there is a gap between theory and practice.\\
\textbf{(iii) Overparameterization and scale bias}. Reparameterizing $\beta=A\gamma$ expands the design to $XA\in\mathbb R^{n\times L}$, creating
severe collinearity and making subsequent LASSO unstable. Because $XA$ cannot be standardized without violating $\beta=A\gamma$, synthetic predictors of small groups face a disproportionally large penalty and tend to be under-selected---a pattern we observe in our microbiome data analysis (Section \ref{sect.realdata}).

In contrast, our method addresses (i)-(iii) by avoiding overparameterization and only penalizing the variability within the candidate aggregating sets implied by the tree. Thus, the selected equi-sparsity structure always coheres with the tree, affording meaningful interpretation. We carefully analyzing the unpenalized subspace and quantify its influence on the prediction error, revealing the theoretical impact of tree structures.

%Paragraph: Our method is inspired by \cite{molstad2023multiresolution}; Write about the philosophy.

%\textcolor{red}{also mention that they only developed theory for uniform weights.}
\vspace{-10pt}
\subsection{Connection to direct regularization methods}\label{sect.method.conn.regularize}
Equi–sparsity can also be achieved by pairwise fusion, e.g., using the fused LASSO
\citep{tibshirani2005sparsity}, which minimizes a loss plus
$\sum_{i<j} w_{ij}\lvert\beta_{i}-\beta_{j}\rvert$.
Related types of pairwise fusion include the 1-D segmented version
\citep{tibshirani2005sparsity}, graph-guided fused LASSO
\citep{chen2010graph}, OSCAR’s $L_\infty$ variant
\citep{bondell2008simultaneous}, and several adaptively weighted variants
\citep{ke2015homogeneity,tang2016fused,chu2021adaptive}. These fusion methods are further related to the topic of convex clustering \citep{chi2025convex}. These methods, however, lack a built-in mechanism for enforcing tree-coherence in our motivating context. Moreover, their theoretical guarantees often require an accurate pilot estimator.

Also worth noting is that our penalty generalizes fusion: for any pair
$\{\beta_{1},\beta_{2}\}$,
\(
  \lvert\beta_{1}-\beta_{2}\rvert
  =\sqrt{2}\,\bigl\|D_{2}(\beta_{1},\beta_{2})^{\!\top}\bigr\|_{2},
\)
so the fused LASSO term is the $a_\ell=2$ instance of our group penalty.
Allowing $a_\ell>2$ enforces group-wise fusion, naturally
accommodates the tree structure and also makes the approach widely applicable to other aggregation tasks.

\vspace{-20pt}
\section{Computation}\label{sect.computation}
\vspace{-10pt}
\subsection{Overview}
\vspace{-10pt}
The optimization problem \eqref{beta-hat} combines a smooth loss \(g\) and the nonsmooth but convex tree penalty
\(\lambda\Omega_{\mathcal T,w}\).
For linear and logistic regression, we use the usual quadratic and cross-entropy losses
\vspace{-10pt}
\begin{align*}
    g(\beta)=
    \begin{cases}
        \frac{1}{2n}\sum_{i=1}^n(y_i-X_i^T\beta)^2 &\text{for linear regression,}\\
        \frac{1}{n}\sum_{i=1}^n\left(\log(1+\exp\{X_i^T\beta\})-y_iX_i^T\beta\right) & \text{for logistic regression.}
    \end{cases}
\end{align*}
In both cases, $g$ is convex and differentiable with Lipschitz-continuous gradient. Thus, we can apply the accelerated proximal gradient descent algorithm \citep{nesterov1983method,parikh2014proximal}, with details provided in the Supplementary Materials \ref{appendix.compute}. The algorithm iteratively updates the optimization variable $\beta$ through the proximal operator of $\lambda\Omega_{\mathcal T,w}$. Specifically, for a step size $\tau > 0$, the $(t+1)$-th iterate is given by
\begin{align}
    \beta^{(t+1)}&=\mathop{\rm argmin}_{\beta \in \mathbb{R}^p} \left\{\frac{1}{2\tau}\left\|\beta-\left[\beta^{(t)}-\tau\nabla g\left(\beta^{(t)}\right)\right]\right\|_2^2+\lambda\Omega_{\mathcal T,w}(\beta)\right\}\notag\\
    &=:\text{prox}_{\tau\lambda \Omega_{\mathcal{T},w}}\{\beta^{(t)}- \tau \nabla g(\beta^{(t)})\}.\label{eq.def.prox}
\end{align}

Thus, the crux of computing \eqref{beta-hat} is evaluating \eqref{eq.def.prox}. %, which decreases the Moreau envelop of the penalty $\Omega_{\mathcal T,w}$. 
With overlapped penalty terms in general, the proximal operator usually requires a numerical solution. Remarkably, in our setting, we show that it can be evaluated using a simple, non-iterative one-pass algorithm. 

Code implementing our method, along with code to reproduce all the results in this article, is available at \url{https://github.com/JinwenFu001/direct_feature_aggregation}.

\vspace{-20pt}
\subsection{Proximal operator for $\Omega_{\mathcal{T},w}$}
\vspace{-10pt}
The proximal mapping\vspace{-20pt}
\begin{equation}
  \vspace{-15pt}
\operatorname{prox}_{\lambda\Omega_{\mathcal{T},w}}(\eta)
=\mathop{\rm argmin}\limits_{\beta}\Bigl\{\tfrac12\|\beta-\eta\|_2^2
      +\lambda\Omega_{\mathcal{T},w}(\beta)\Bigr\}\label{eq:prox}
\end{equation}
is uniquely defined because the objective in \eqref{eq:prox} is strongly convex.  
Although the tree‐structured penalty precludes a closed-form solution in general,
the map can be evaluated exactly by a ``bottom-up'' sequence of shrinkage updates, each of which has a simple closed form. To illustrate this algorithm, define\vspace{-15pt}
\begin{gather}
  \vspace{-15pt}
\tilde\beta_0
=\mathop{\rm argmin}\limits_{\beta}\Bigl\{
        \tfrac12\|\beta-\eta\|_2^2
        +\lambda\!\!\sum_{\ell\in\mathcal I^\star}\!\!
        w_\ell\|D_{a_\ell}\beta_{\mathcal{A}_\ell}\|_2
      \Bigr\}, \label{eq:prox-sub}\\
\tilde\beta
=\mathop{\rm argmin}\limits_{\beta}\Bigl\{
        \tfrac12\|\beta-\eta\|_2^2
        +\lambda w_{\ell_0}\|D_{a_{\ell_0}}\beta_{\mathcal{A}_{\ell_0}}\|_2
        +\lambda\!\!\sum_{\ell\in\mathcal I^\star}\!\!
        w_\ell\|D_{a_\ell}\beta_{\mathcal{A}_\ell}\|_2
      \Bigr\}, \label{eq:prox-full}
\end{gather}
where $\{\mathcal{A}_\ell\}_{\ell\in\mathcal I^*}$ and $\mathcal{A}_{\ell_0}$ correspond to internal nodes $\{b_\ell\}_{\ell\in\mathcal I^*}\subset\mathcal{T}$ and $b_{\ell_0}$.
Going from \eqref{eq:prox-sub} to \eqref{eq:prox-full}, we add the penalty on group $\mathcal{A}_{\ell_0}$. In the following, we assume that corresponding internal node $b_{\ell_0}$ is not a descendant of any node in
$\{b_\ell\}_{\ell\in\mathcal I^\star}$. In this case, the following theorem shows that $\tilde\beta$ can be expressed in closed form using $\tilde\beta_0$.

\begin{thm}[Closed-form update]\label{thm:closed}
If $\tilde\beta_0$ solves \eqref{eq:prox-sub}, and $b_{\ell_0}$ is not a descendant of any node in $\{b_\ell\}_{\ell\in\mathcal{I^*}}$, then $\tilde\beta$ in
\eqref{eq:prox-full} is also given by the solution of the non-overlapping problem
\vspace{-15pt}
\[
\tilde\beta=\mathop{\rm argmin}_{\beta}\Bigl\{
        \tfrac12\|\beta-\tilde\beta_0\|_2^2
        +\lambda w_{\ell_0}\|D_{a_{\ell_0}}\beta_{\mathcal{A}_{\ell_0}}\|_2
      \Bigr\}.
\]
Define the penalty-variability ratio $\rho_{\ell_0}=\lambda w_{\ell_0}/\|D_{a_{\ell_0}}[\tilde\beta_{0}]_{\mathcal{A}_{\ell_0}}\|_2$ and the within group average $\bar\beta_{\ell_0}=a_{\ell_0}^{-1}\mathbf 1_{a_{\ell_0}}^\top[\tilde\beta_{0}]_{\mathcal{A}_{\ell_0}}$, the closed-form solution is
\begin{align*}
[\tilde\beta]_{\mathcal{A}_{\ell_0}}&=
\begin{cases}
\bar\beta_{\ell_0}\mathbf 1_{a_{\ell_0}},
& \rho_{\ell_0}\ge 1\\[4pt]
\rho_{\ell_0}\bar\beta_{\ell_0}\mathbf 1_{a_{\ell_0}}
+(1-\rho_{\ell_0})[\tilde\beta_{0}]_{\mathcal{A}_{\ell_0}},
& \rho_{\ell_0}<1
\end{cases} ~~~\text{and }~~
[\tilde\beta]_{\mathcal{A}^c_{\ell_0}}=[\tilde\beta_{0}]_{\mathcal{A}^c_{\ell_0}},
\end{align*}
where $\mathcal{A}_{\ell_0}^c = [p] \setminus \mathcal{A}_{\ell_0}$ , and $[\tilde\beta]_{\mathcal{A}_{\ell_0}}$ is the subvector of $\tilde\beta$ indexed by $\mathcal{A}_{\ell_0}$.
\end{thm}

\begin{figure}[t]
\centering
\begin{subfigure}{0.32\textwidth}\centering
\includegraphics[width=\linewidth]{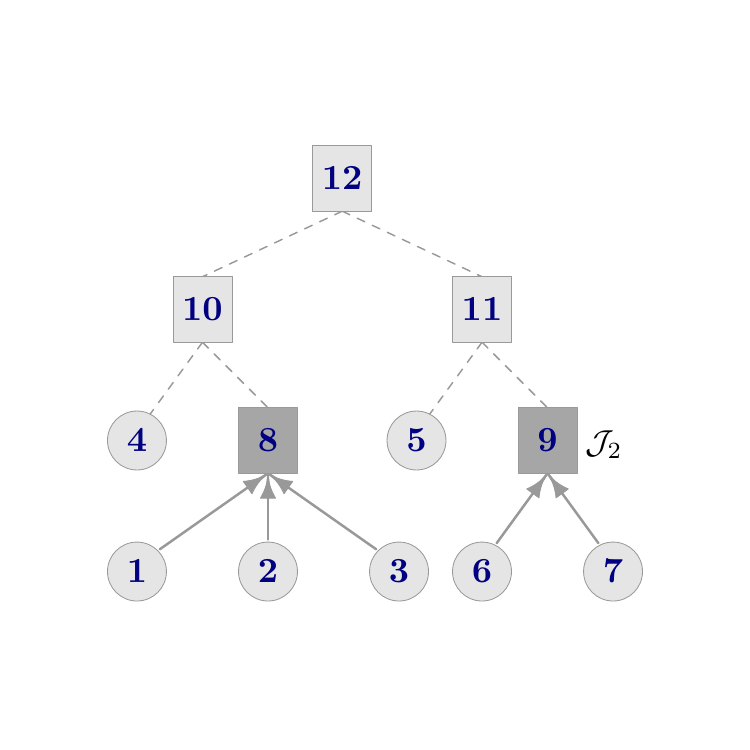}
%\caption{}\label{fig:prox.1}
\end{subfigure}\hfill
\begin{subfigure}{0.32\textwidth}\centering
\includegraphics[width=\linewidth]{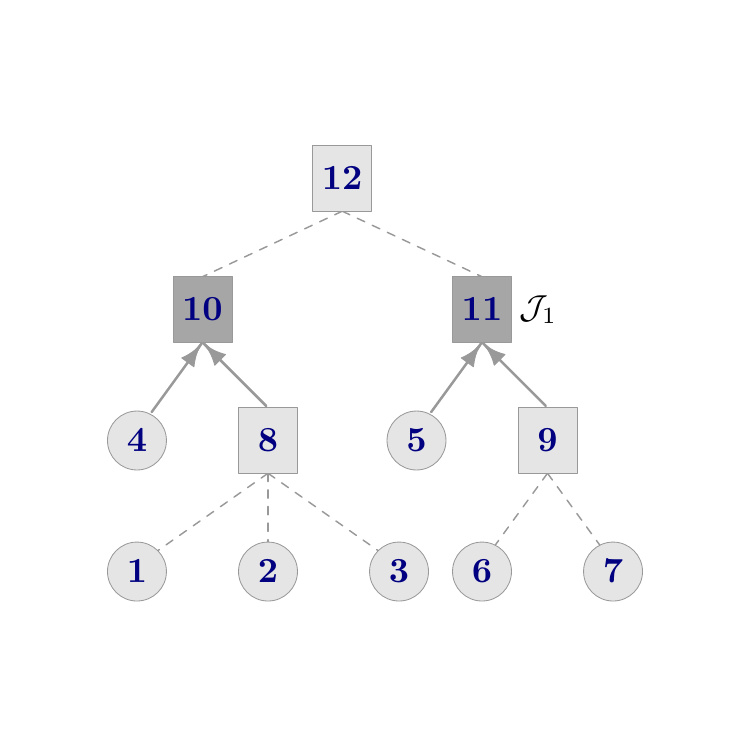}
%\caption{}\label{fig:prox.2}
\end{subfigure}
\begin{subfigure}{0.32\textwidth}\centering
\includegraphics[width=\linewidth]{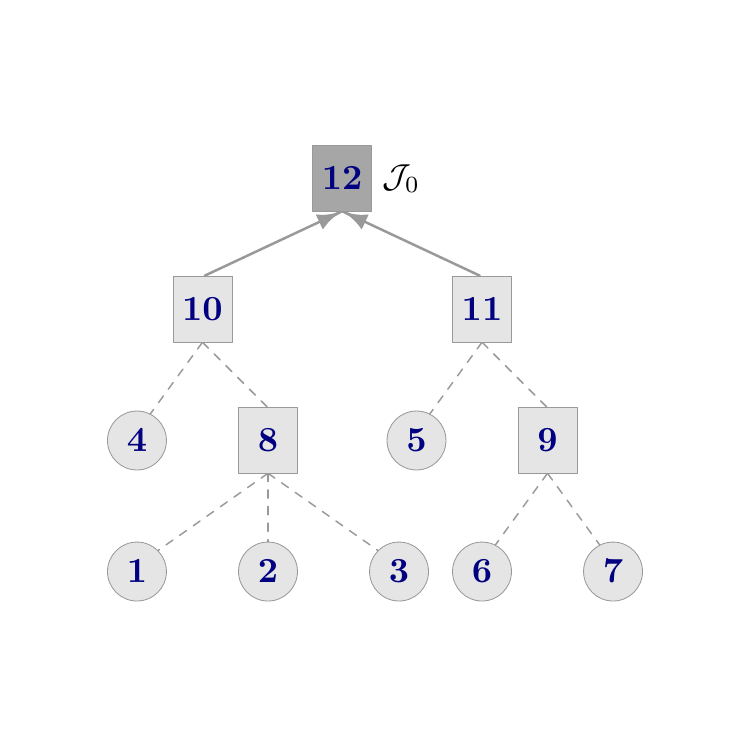}
%\caption{}\label{fig:prox.3}
\end{subfigure}
\caption{An example of the bottom-up evaluation for proximal operator. The newly added internal nodes $\mathcal J_0,~\mathcal J_1$ and $\mathcal J_2$ in each step are colored in grey.}\label{fig:prox}
\end{figure}

Applying the result of Theorem $\ref{thm:closed}$, we see that the proximal operator \eqref{eq:prox} can be computed using a bottom-up, one-pass algorithm wherein we traverse the tree from leaves to root and sequentially add penalty terms. To start with, we group the latent nodes by depth: $(\mathcal{J}_h,\ldots,\mathcal{J}_0)$, where $\mathcal{J}_0$ contains only the roots. We describe the procedure in Algorithm \ref{alg:prox}, which is also visualized in Figure \ref{fig:prox}.

\begin{algorithm}[H]
\caption{One-pass evaluation of $\operatorname{prox}_{\lambda\Omega_{\mathcal{T},w}}(\eta)$}
\label{alg:prox}
\begin{algorithmic}[1]
\Procedure{Prox}{$\mathcal{T},\eta,w,\lambda$}
\State\textbf{initialize  } $\hat\beta^{(0)}
       \gets\mathop{\rm argmin}_{\beta}\bigl\{\tfrac12\|\beta-\eta\|_2^2
             +\lambda\!\!\sum_{\ell\in\mathcal{J}_h}\!\!
             w_\ell\|D_{a_\ell}\beta_{\mathcal{A}_\ell}\|_2\bigr\}$.
\For{$i=1$ \textbf{to} $h$}
    \State $\hat\beta^{(i)}
           \gets\mathop{\rm argmin}_{\beta}\bigl\{\tfrac12\|\beta-\hat\beta^{(i-1)}\|_2^2
                 +\lambda\!\!\sum_{\ell\in\mathcal{J}_{h-i}}\!\!
                 w_\ell\|D_{a_\ell}\beta_{\mathcal{A}_\ell}\|_2\bigr\}$ 
\EndFor
\State \Return $\hat\beta^{(h)}$
\EndProcedure
\end{algorithmic}
\end{algorithm}
\vspace{-15pt}
The solution reveals several interesting properties of the penalty. First of all, each update doesn't change the summation of values across $\beta$, but only redistributes mass inside the group through a linear
shrinkage toward $\mathbf 1_{a_{\ell_0}}$. From the step \eqref{eq:prox-sub} to \eqref{eq:prox-full}, the summation is preserved within every newly penalized group $\mathcal{A}_{\ell_0}$,\vspace{-15pt}
\[
\vspace{-15pt}
\mathbf 1_{a_{\ell_0}}^\top[\tilde\beta]_{\mathcal{A}_{\ell_0}}
   =\mathbf 1_{a_{\ell_0}}^\top[\tilde\beta_0]_{\mathcal{A}_{\ell_0}},
\qquad
\mathbf 1_{p-a_{\ell_0}}^\top[\tilde\beta]_{\mathcal{A}^c_{\ell_0}}
   =\mathbf 1_{p-a_{\ell_0}}^\top[\tilde\beta_0]_{\mathcal{A}^c_{\ell_0}}.
\]
Hence, the proximal map evaluated by Algorithm \ref{alg:prox} preserves the property $\mathbf 1_p^\top\tilde\beta^{(h)}=\mathbf 1_p^\top\eta$. This illustrates how our penalty performs shrinkage towards aggregated models, not towards the origin. It is this mechanism that leads our estimates to be less biased towards the origin than those that use the overparameterize-regularize approach. 

\vspace{-15pt}
\section{Statistical properties}\label{sect.theory}
\vspace{-10pt}
\subsection{Preliminaries}
\vspace{-10pt}
In this section, we study the prediction and estimation consistency of the proposed method. Given loss function $g$, we study the distance between regularized estimator\vspace{-10pt}
\[
\vspace{-10pt}
\hat\beta\in\mathop{\rm argmin}_{\beta \in \mathbb{R}^p} \left\{g(\beta)+\lambda\Omega_{\mathcal{T},w}(\beta)\right\}
\]
and the parsimonious target $\beta^*$. We focus on the regression and classification settings. For regression, we consider $g(\beta)=\frac{1}{2n}\|y - X\beta\|^2_2$, i.e., least squares loss. In this context, a finite sample bound is provided for the predictive error $\|X\hat\beta-X\beta^*\|^2_2$ in subsection \ref{subsect.pred.error.bound}, illustrating the impact of the tree structure $\mathcal{T}$ on prediction accuracy. Subsection \ref{subsect.est.error.bound1} further improves the rate under an additional condition for the design matrix $X$, which also allows us to establish an error bound for $\|\hat\beta-\beta^*\|_2$. For classification tasks, we use the logistic regression model and take $g(\beta)=n^{-1}\sum_{i=1}^n[\log\{1+\exp(X_i^T\beta)\}-y_iX_i^T\beta]$. Estimation consistency under this scenario is studied in the third subsection \ref{subsect.est.error.bound2}. 

Throughout the section, we use default choice of weight $w_\ell= a_\ell^{-1/2}$, so that the penalty can be expressed
$
\Omega_{\mathcal{T},w}(\beta)=\sum_{\ell\in\mathcal{I}}\frac{1}{\sqrt{a_\ell}}\|D_{a_\ell}\beta_{\mathcal{A}_\ell}\|_2.
$
To make the analysis easier, we define the matrix $M(\mathcal{A}_\ell)\in\mathbb R^{p\times p}$ such that its submatrix with rows and columns indexed by $\mathcal{A}_\ell$ being equal to $D_{a_\ell}$ with the other entries being zero. As such, $[M(\mathcal{A}_\ell)\beta]_{\mathcal{A}_\ell}=D_{a_\ell}\beta_{\mathcal{A}_\ell}$, and the penalty can be equivalently written as
$
\Omega_{\mathcal{T},w}(\beta)=\sum_{\ell\in[L]}\frac{1}{\sqrt{a_\ell}}\|M(\mathcal{A}_\ell)\beta\|_2.
$ Note that when $\mathcal A_\ell$ is a singleton representing a leaf node, $a_\ell=1$ means $D_{a_\ell}=0$ and $M(\mathcal A_\ell)=(\mathbf 0)_{p\times p}$ is a zero matrix, then the corresponding penalty term is zero.

In contrast to LASSO and other penalties that are induced from a norm and regularize $\beta$ towards the origin, the proposed penalty $\Omega_{\mathcal T,\omega}(\beta)$ is a seminorm and shrinks $\hat\beta$ towards a subspace. The difference turns out to play a nontrivial role in the analysis and requires more delicate treatment. Therefore, we first define the kernel of the penalty term $\Omega_{\mathcal T,\omega}(\beta)$.

\begin{definition}[Kernel of $\Omega_{\mathcal{T},w}$]
    We define the kernel of the proposed penalty as the kernel of the corresponding seminorm, denoted by $\Ker(\Omega_{\mathcal{T}},\omega)=\{\nu\in\mathbb R^p: \Omega_{\mathcal{T},\omega}(\nu)=0\}$. The kernel nodes set, denoted by $\mathcal N=\{\ell\in[L]:\nexists\ \tilde \ell\neq\ell$ s.t. $\mathcal A_{\ell}\subset  \mathcal A_{\tilde\ell}\}$, is the set of nodes in the tree that have no ancestors. The kernel node set is an aggregating set and represents the highest level of aggregation for the tree $\mathcal T$.
\end{definition}\label{def.kernel}
The kernel consists of the directions along which $\Omega_{\mathcal T,\omega}(\beta)$ stays unchanged when $\beta$ is perturbed. Intuitively, we can view the kernel node set as the roots of the tree (or the forest) structure. We take the example tree in Figure \ref{fig.eg.tree} for illustration. The penalty 
$
\Omega_{\mathcal{T},w}(\beta)= 7^{-1/2} \|M(\mathcal{A}_{{11}}) \beta \|_2+ 3^{-1/2}\|M(\mathcal{A}_{9})\beta\|_2 +2^{-1/2}\|M(\mathcal{A}_{8})\beta \|_2 +2^{-1/2}\| M(\mathcal{A}_{{10}})\beta\|_2
$ 
shrinks parameter $\beta$ towards the subspace $\Ker(\Omega_{\mathcal{T}},\omega)=\mathop{\rm span}\{\mathbf 1_7\}$, which corresponds to the kernel node set $\mathcal{N}=\{11\}$, a singleton of the top layer node. We further study the property of the kernel in the Supplementary Materials \ref{appendix.dimension}.

We decompose the internal nodes $\mathcal{I}$ into two parts that reflect the structure of true $\beta^*$. Let $\mathcal I_1$ be the set of latent nodes $\ell_1$ from the tree such that the leaf nodes of the corresponding subtree should not be fully aggregated, i.e. $\|M({\mathcal A_{\ell_1}})\beta^*\|_2\neq 0$; Similarly, $\mathcal I_0$ contains the latent nodes such that $\|M({\mathcal A_{\ell_0}})\beta^*\|_2= 0$.

%With the above definitions, we can now study the regularized estimator.
\vspace{-10pt}
\subsection{Condition-free prediction error bound for linear regression}\label{subsect.pred.error.bound}
 We first consider the linear regression setting with data generating model
 \(y = X\beta^* + \epsilon,\)
 and treat the design matrix as deterministic to study the predictive error \(\|X\hat{\beta} - X\beta^*\|_2^2\). In this section, we impose only minimal regularity conditions on \(X\). Thus, we call our error bound a ``condition-free'' error bound. We will make the following assumptions.

 \begin{assump}[Sub-Gaussian Noise]\label{assump.model}
    The elements in noise \(\epsilon\) are independent and mean zero \(\sigma\)-sub-Gaussian random variables.
\end{assump}

\begin{assump}[Parameter Structure]\label{assump.true.beta}
    The parameter \(\beta^*\) takes the form
    $
      \beta^* = \bigl(\beta^*_1 \otimes \mathbf{1}_{d_1}^T,\;\dots,\;\beta^*_K \otimes \mathbf{1}_{d_K}^T\bigr)^T,$
    consisting of \(K\) groups with size \(d_1,\cdots, d_K\), and sharing the common effect value \(\beta^*_k\) within each group. 
\end{assump}
 %\begin{assump}
 %    There exists an aggregating set $\mathcal B^*\in\mathcal L$ for the tree $\mathcal T$ that corresponds to the group structure of $\beta^*$ in Assumption \ref{assump.true.beta}.\label{assump.tree}
 %\end{assump}
 \begin{assump}[Scaling]
     There exists a constant $C$, such that $\max_{\ell\in\mathcal I}\vvvert{\frac{X_{\cdot \mathcal A_\ell}}{\sqrt{n}}}\le C < \infty$, where $\vvvert{\cdot}$ is the spectral norm of its matrix-valued input. \label{assump.scaling}
 \end{assump}

Assumptions~\ref{assump.model} and~\ref{assump.true.beta} specify the true linear model and its equi-sparsity structure. Assumption~\ref{assump.scaling} enforces the standard group-scaling condition on the design matrix, in line with group LASSO \citep{yuan2006model}, requiring the largest spectral norm across group-specific design matrices to be bounded. The assumption, despite being imposed on the design matrix, can be easily satisfied by an arbitrary matrix $X$ through scaling. %Under those assumptions, we state our first major result, which is proved in appendix \ref{proof.thm.pred.bound1}.

\begin{thm}
    Under the Assumption \ref{assump.model}, \ref{assump.true.beta} and \ref{assump.scaling}, if the user-specified tree $\mathcal T$ is correct in that it encodes an aggregating set $\mathcal B^*\in[L]$ corresponding to the group structure of $\beta^*$ in Assumption \ref{assump.true.beta}, and we take the penalty parameter to be\vspace{-10pt}
    \begin{equation}
      \vspace{-10pt}
        \lambda= \frac{4\sqrt{2}\sigma C}{\sqrt n}\Theta(\mathcal T)\sqrt{\log (2p)+\log(|\mathcal I|)}~~
    \text{ where }~~~\Theta(\mathcal T)= \sqrt{\max_{\ell\in\mathcal{I}}a_\ell}\left(1+\sqrt{\frac{\max_{\ell\in\mathcal I}{a_\ell}}{\log (|\mathcal{I}|)}}\right),\notag
    \end{equation}
     then with probability at least $1-2/p$,\vspace{-10pt}
    \begin{equation}
    \begin{split}
        \frac{1}{n}\|X(\hat\beta-\beta^*)\|_2^2&\le \frac{24\sqrt{2}\sigma C}{\sqrt n}\Theta(\mathcal T)\sqrt{\log (2p)+\log(|\mathcal I|)} \left(\sum_{\ell_1\in\mathcal I_1}\frac{1}{\sqrt{ a_{\ell_1}}}\|M(\mathcal{A}_{\ell_1})\beta^*\|_2\right) \\
     & \quad\quad + \frac{32\sigma^2\{\log(2p)+\min(|\mathcal{N}|,n)\log(5)\}}{n},\vspace{-10pt}
    \end{split}
    \end{equation}
    where $\mathcal{I}_1$ denotes the set of nodes in the tree that should not be aggregated.\label{thm.pred.bound1}
\end{thm}
\begin{remark}
    The quantity $\Theta(\mathcal T)$ is defined regarding the given tree. Intuitively, $\Theta(\mathcal T)$ grows with the size of the largest group and the number of nodes indicated by the tree. It quantifies the selection complexity of the group-wise penalty analogous to the analysis of group LASSO \citep[Chapter 9]{wainwright2019high}. 
\end{remark}

Unlike typical high-dimensional analysis, the result in Theorem~\ref{thm.pred.bound1} features two additive terms. The second term arises from the penalty's null space whose cardinality $|\mathcal{N}|$, understood as the number of roots in the tree (or the forest) structure, is typically much smaller than $n$ and $p$. Hence, the second term is square order for $\sqrt{\log(p)/n}$ and usually becomes negligible with large $n$. Consequently, the first term is more influential in determining the primary error rate: its main driver is
\[
  \sum_{\ell_1 \in \mathcal{I}_1} \frac{1}{\sqrt{a_{\ell_1}}} \, \bigl\|M(\mathcal{A}_{\ell_1})\beta^*\bigr\|_2=\sum_{\ell_1 \in \mathcal{I}_1} \frac{1}{\sqrt{a_{\ell_1}}} \, \bigl\|D_{\mathcal{A}_{\ell_1}}\beta^*_{\mathcal{A}_{\ell_1}}\bigr\|_2,
\]
where \(\mathcal{I}_1\) is the set of groups in the tree that in truth, are not aggregated. This term illustrates the price we pay searching for the correct aggregating sets using the tree. Hence, to mitigate this price, the theorem suggests pruning high-level branches that are unlikely to reflect correct aggregations before analyzing the data. The same pattern is observed in simulation studies.

While the other coefficients in the first term remain complicated in representing the tree configuration, we further give two examples of specific tree structures, allowing the number of true groups $k$ to scale differently with growing dimension $p$. With a finite sample $n$ and under the specific tree structures, we can simplify the above result to have similar form with the slow rate bound of LASSO, and show theoretical comparison of the proposed method with the method in \cite{yan2021rare} or other regularization.

\subsubsection{Direct Comparison with RARE}

\begin{example}[\textbf{Expanding parameter model}]\label{exp:expand-tree}
Consider a setting where the number of true groups in \(\beta^*\) grows linearly with dimension \(p\), so neither $K$ nor $p$ is considered constant in the analysis. In this situation, the tree is made up of \(t\) copies of the same subtree, each containing \(p_s\) parameters within the \(K_s\) groups. Taking $p_s$ and $K_s$ as constants, the total dimension \(p = t\,p_s\) and the total number of groups \(K=tK_s\) increase with the number of subtree copies $t$. The following corollary shows that, under this scenario, the predictive error rate is at least as good as LASSO's “slow rate,” matching the rate of \cite{yan2021rare}.
\begin{corol}
Under the expanding tree model in Example \ref{exp:expand-tree} and the assumptions of Theorem \ref{thm.pred.bound1}, with probability at least $1-2/p$, we have 
$
n^{-1}\,\|X(\hat\beta-\beta^*)\|_2^2 
\;\lesssim\; 
\sqrt{n^{-1}\log (p)} \,\|\beta^*\|_1,
$
where `\(\lesssim\)` indicates that we ignore constant factors and higher-order terms.\label{corol.tree1}
\end{corol}
\end{example}
The previous example demonstrates that our rate is no worse than \citet{yan2021rare} for certain tree structures.  The following corollary shows that under some other tree models, we are able to achieve a faster rate than the slow rate bound of LASSO and the method of \cite{yan2021rare}.

\begin{example}[\textbf{Fixed parameter model}]\label{exp:fixed-tree}
Suppose we fix the number of true groups in \(\beta^*\) to a constant \(K\), while allowing the dimension \(p\) to grow. We assume that the portion of the tree structure above the latent nodes representing these true groups, denoted by \(\mathcal{I}_1\), remains unchanged as $p$ grows. 
\begin{corol}
Suppose that there exists a constant $\rho_C$ such that \(\rho = \max_{\ell \in \mathcal{B}^*} a_{\ell}/\min_{\ell \in \mathcal{B}^*} a_{\ell} \le \rho_C\), where \(\mathcal{B}^*\) is the aggregating set in Definition~\ref{def.agg.set} and \(\rho_C\) is a constant. Then, under the fixed tree model in Example \ref{exp:fixed-tree} the assumptions of Theorem \ref{thm.pred.bound1},  with probability at least \(1 - 2/p\),
$
n^{-1}\|X (\hat\beta-\beta^*)\|_2^2
\;\lesssim\;
\left[n^{-1/2} + \sqrt{(np)^{-1} \log (p)} \,\right]\|\beta^*\|_1.
$\label{corol.tree2}
\end{corol}
\end{example}
The result of the preceding corollary suggests an improvement by a factor of $\{\log(p)\}^{-1/2}$ over the method of \citet{yan2021rare}, which is further verified by simulation results in Section \ref{sect.experiment.1}. %The proofs of Corollary \ref{corol.tree1} and \ref{corol.tree2} can be found in Appendix \ref{proof.corol.trees}.

\vspace{-20pt}
\subsection{Estimation error bound for linear regression}\label{subsect.est.error.bound1}
Next, we study the estimation error \(\|\hat\beta - \beta\|^2\) under an additional condition on the design matrix. We show that, with this extra condition, the predictive error bound can be further improved compared to Theorem~\ref{thm.pred.bound1}. It also provides an alternative way to understand the role of the tree in our method.

\begin{assump}[$\mathcal{T}$-restricted eigenvalue condition]\label{assump.RE}
    Let 
    \[
    S_{\mathcal{T}} 
    \;=\; 
    \Bigl\{
      \Delta : \;\sum_{\ell_0 \in \mathcal{I}_0} \frac{1}{\sqrt{a_{\ell_0}}}\,\|M({\mathcal{A}_{\ell_0}})\Delta\|_2
      \;\le\;
      3\sum_{\ell_1 \in \mathcal{I}_1} \frac{1}{\sqrt{a_{\ell_1}}}\,\|M({\mathcal{A}_{\ell_1}})\Delta\|_2
      \;+\;
      \frac{\sqrt{|\mathcal N|}}{8\Theta(\mathcal{T})}\|\Delta\|_2
    \Bigr\}.
    \]
    There exists a constant \(\kappa > 0\) such that, for every \(\Delta \in S_{\mathcal{T}}\),
    $
    \frac{1}{n} \|X \Delta\|_2^2 \;\ge\; \kappa \,\|\Delta\|_2^2.
    $
\end{assump}

The restricted eigenvalue condition is an assumption on the design matrix \(X\), requiring it be sufficiently well-conditioned in directions  \(\Delta \in S_{\mathcal{T}}\). The restricted eigenvalue $\kappa$ is partly determined by the “volume” of \(S_{\mathcal{T}}\), which, in turn, is determined by the tree structure \(\mathcal{T}\). Intuitively, removing latent nodes in \(\mathcal{I}_0\) decreases 
\(\sum_{\ell_0 \in \mathcal{I}_0} \frac{1}{\sqrt{a_{\ell_0}}}\|D_{\mathcal{A}_{\ell_0}}\Delta\|_2\),
thus increasing the volume of \(S_{\mathcal{T}}\) and leading to a smaller \(\kappa\). Conversely, pruning some nodes in \(\mathcal{I}_1\) might reduce the size of \(S_{\mathcal{T}}\) and yield a larger \(\kappa\). The last term in $S(\mathcal{T})$ has a relatively small impact considering that $\sqrt{|\mathcal{N}|}$ is typically much smaller than $\Theta(\mathcal{T})$. We now state our main result under this additional assumption.

\begin{thm}\label{thm.est.bound}
    Under Assumptions~\ref{assump.model}, \ref{assump.true.beta}, \ref{assump.scaling}, and the restricted eigenvalue condition~\ref{assump.RE}, if the tuning parameter $\lambda$ is chosen as in Theorem \ref{thm.pred.bound1}, then with probability at least \(1 - 2/p\), 
    \begin{equation}
        \|\hat\beta - \beta^*\|_2
        \;\le\;
        \frac{6\sqrt{2}\sigma C}{\kappa\sqrt n}\Theta(\mathcal T)\sqrt{{\log (2p)}+{\log(|\mathcal I|)}}\left(\sum_{\ell_1\in\mathcal I_1}\frac{1}{\sqrt{a_{\ell_1}}}\right)+\frac{\sigma C}{\kappa}\sqrt{\frac{|\mathcal N|\log (2p |\mathcal{N}|)}{2n}}
    \end{equation}
    and also
    \begin{equation}\label{eqn.est.pred}
    \begin{split}
        \frac{1}{\sqrt{n}}\|X(\hat\beta - \beta^*)\|_2
        \;&\le\;
        \frac{12\sqrt{2}\sigma C}{\sqrt{\kappa n}}\Theta(\mathcal T)\sqrt{{\log (2p)}+{2\log(|\mathcal I|)}}\left(\sum_{\ell_1\in\mathcal I_1}\frac{1}{\sqrt{a_{\ell_1}}}\right)\\
        &+\frac{\sigma C}{\sqrt{\kappa}}\sqrt{\frac{2|\mathcal N|\log (2p |\mathcal N|)}{n}}.
    \end{split}
    \end{equation}
\end{thm}
%See Appendix~\ref{proof.thm.est.bound} for the proof of Theorem \ref{thm.est.bound}.

Theorem~\ref{thm.est.bound} highlights how a suitable choice of tree structure can improve error bounds. Because the predictive error term is inflated by a factor of \(1/\sqrt{\kappa}\), it is advantageous to preserve more nodes in \(\mathcal{I}_0\) and prune additional nodes in \(\mathcal{I}_1\). In conjunction with Theorem~\ref{thm.pred.bound1}, our results suggest that the tree should prioritize aggregating groups near the leaf nodes, while being more conservative about preserving branches near the root. This conclusion is further supported by our simulation studies.

\subsection{Estimation error bound for logistic regression}\label{subsect.est.error.bound2}

In the classification context, we consider generalized linear model with logit link. Specifically, the following logistic model is assumed for independent binary outcomes $\{y_i\}_{i=1}^n$,
$P(y_i=1)/P(y_i=0) ={\rm exp}(X_i^T\beta^*).$
The estimator we consider is
$\hat\beta\in\mathop{\rm argmin}_{\beta\in\mathbb{R}^p} \left\{g(\beta)+\lambda\Omega_{\mathcal T,w}(\beta)\right\},$
though here, $g(\beta)=\frac{1}{n}\sum_{i=1}^n[\log\{1+\exp(X_i^T\beta)\}-y_iX_i^T\beta ]$ is the cross-entropy loss. To study the estimation error $\|\hat\beta-\beta^*\|^2_2$, we further refine the curvature of the loss function on a restricted set around the true value $\beta^*$.
\begin{assump}[$\mathcal{T}$-restricted strong convexity]
\label{assump.RSC}
    Let $S_{\mathcal{T}}$ be the same set as defined in Assumption \ref{assump.RE}. There exists a constant \(\kappa > 0\) such that, for every \(\Delta \in S_{\mathcal{T}}\),
    $g(\beta^*+\Delta)-g(\beta^*)\ge\nabla g(\beta^*)^T\Delta+\frac{\kappa}{2}\|\Delta\|_2^2.
    $
\end{assump}
Replacing Assumptions \ref{assump.model} and \ref{assump.RE} with Assumption \ref{assump.RSC}, we have the following theorem describing the finite sample error bound of the proposed estimator.

\begin{thm}\label{thm.logit.est.bound}
    Under Assumptions~\ref{assump.true.beta}, \ref{assump.scaling}, and \ref{assump.RSC}, 
    if we take 
   $\lambda = \frac{2\sqrt{2}C}{\sqrt n}\Theta(\mathcal T)\sqrt{{\log (2p)}+{\log|\mathcal I|}},$
    then with probability at least \(1 - 2/p\), \vspace{-15 pt}
    \begin{align}
        \|\hat\beta - \beta^*\|_2
        \;&\le\;
        \frac{3\sqrt{2}C}{\kappa\sqrt n}\Theta(\mathcal T)\sqrt{{\log (2p)}+{\log|\mathcal I|}}\sum_{\ell_1\in\mathcal I_1}\frac{1}{\sqrt{a_{\ell_1}}}+\frac{ C}{\kappa} \sqrt{\frac{|\mathcal N|\log(2p|\mathcal N|)}{8n}}\vspace{-10pt}
    \end{align}
\end{thm}
%A proof of the preceding result can be found in Appendix \ref{proof.thm.logit.est.bound}. 
It is possible to establish an error bound for the log-odds with a slight modification of the proof, though we do not pursue this here for the sake of space. 

\vspace{-20pt}
\section{Simulation studies} \label{sect.simulation}
%\vspace{-10pt}
% Various simulations are conducted to study the empirical properties of the proposed method. In the first subsection, experiments showcase the validity of our theory and elucidate the impact of the tree structure on prediction accuracy. The second subsection presents the comparison between our methods and several competitors, demonstrating a general advantage across different settings.  %In each setting, we first generate a tree $\mathcal{T}$ with $p$ leaves and make sure it encodes the correct aggregation information for $k$ clusters. The first half of clusters each has $\frac{3p}{2k}$ features, and the remaining half clusters each has $\frac{p}{2k}$ features. 

%We simulate $k$ independent values from Uniform(1.5,2.5) and alternate the signs to obtain true values for cluster effects. The entries of design matrix $X\in \mathbb{R}^{n\times p}$ are simulated independently from Poisson(0.02) distribution.
\vspace{-10pt}
\subsection{Validation of the theory}
\vspace{-10pt}
The theory we established reveals two interesting facts: (i) the structure of the tree largely impacts the prediction and estimation accuracy, as observed in Theorem \ref{thm.pred.bound1} and \ref{thm.est.bound}; (ii) under some specific design of the tree, we are able to achieve a better prediction rate than that of the method of \citet{yan2021rare} (Corollary \ref{corol.tree2}). The two experiments in this subsection show that these conclusions can also be drawn empirically.
\vspace{-10pt}
\subsubsection{Effect of number of internal nodes}\label{sect.experiment.1}
\begin{figure}[t]
  \centering
  \includegraphics[width=\linewidth]{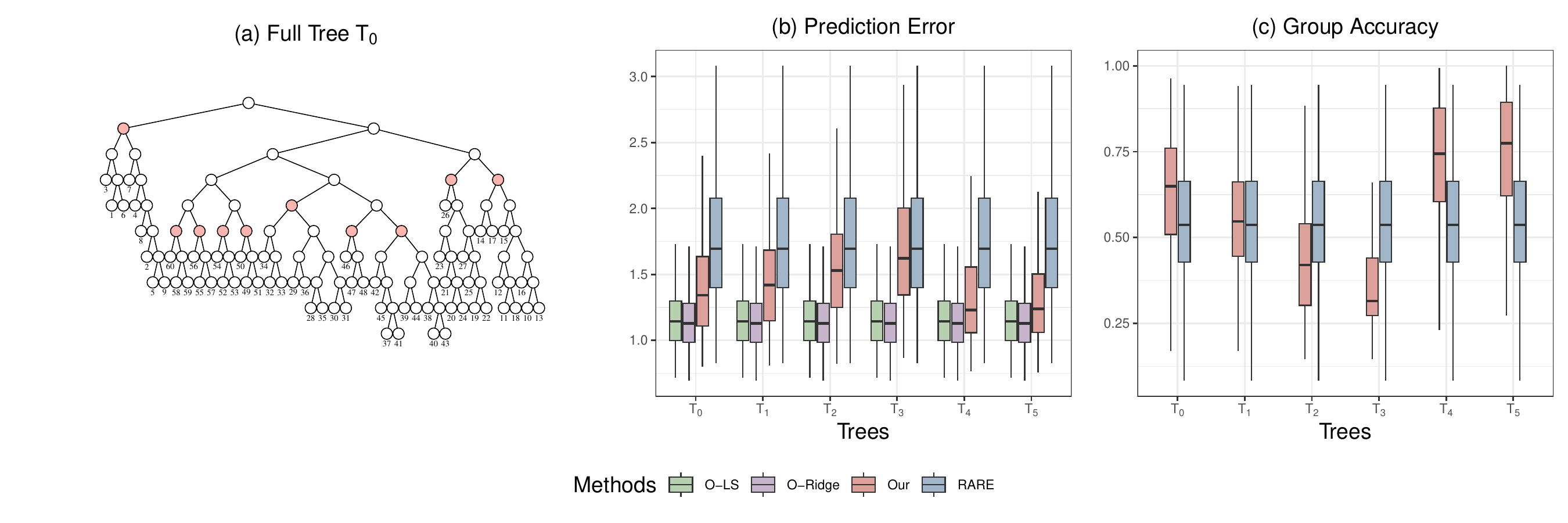}
  \caption{Results of the simulation studies described in Section \ref{sect.experiment.1}. (a) The original tree structure $\mathcal{T}_0$, from which we generate $\mathcal{T}_1,\cdots,\mathcal{T}_5$ by deleting nodes. (b) The boxplot of the prediction error of the four methods under 200 replications. (c) The contrast of group accuracy (adjusted Rand index) of our method and RARE under different tree structures.}
  \label{fig:three_panels}
\end{figure}

\vspace{-10pt}
We start from generating a fixed tree structure, as shown in Figure \ref{fig:three_panels}a. The total dimension $p$ of the predictors is set to be 60, represented by the leaves of the tree. We set the number of true effect groups $K=10$, the corresponding aggregating set is highlighted by those shaded latent nodes in the graph. The first 5 groups each has 9 features; and the remaining 5 groups each has 3 features. For such a given tree, We simulate $K=10$ independent values from Uniform(1.5,2.5) and alternate the signs to obtain the true effect values. The entries of the design matrix $X\in\mathbb R^{n\times 60}$ are generated independently from Poisson(0.02). Then we obtain $y\sim N(X\beta^*,\sigma^2 \mathbf{I}_n)$, where $\sigma^2=\|X\beta^*\|_2^2/(5n)$. We train the model on a training set with size 50 and select the penalty parameter on a validation set with the same sample size. The prediction errors are evaluated on an independent validation set with size 500. We repeat the procedure 200 times and apply the proposed method to the generated datasets. 

The first column in Figure \ref{fig:three_panels}b shows our baseline behavior, using the full tree and compared to the rare feature selection method. Now we modify the tree structure by randomly deleting latent nodes, either above or below the aggregating set. However, we preserve the entire aggregating set so that Assumption \ref{assump.true.beta} still holds. Upon deletion of a node, we connect all of its children directly to its parent. We use $\mathcal T_0$ to denote the full tree, and generate trees $\mathcal T_1,\cdots,\mathcal T_5$ by randomly deleting 10, 20, 30 nodes below (in $\mathcal{I}_0$) and 3, 6 nodes above the aggregating set (in $\mathcal{I}_1$), respectively. See the Supplementary Materials \ref{appendix.tree.plots} for a visualization of the tree structures. Figure \ref{fig:three_panels}b and \ref{fig:three_panels}c summarize the predicted error and adjusted Rand index using the same datasets but different trees. A dramatic increase in the prediction error and decrease in the grouping accuracy are observed when more latent nodes are deleted below the aggregating set, where the corresponding parameters are supposed to be merged. In contrast, deleting nodes above the aggregating set improves the behavior and gives a smaller prediction error and a larger Rand index. It aligns with the theorem that we prefer a tree with as much precise information as possible above the true aggregating set; and as many branches as possible below the aggregating set, to reduce the volume of the restricted set $S_{\mathcal T}$ and obtain a larger strong convexity constant $\kappa$. We also measure the prediction error using oracle least square and oracle ridge as the benchmark performance. They assume we have the oracle group information and directly regress over the correctly aggregated groups. With a proper tree structure such as $\mathcal T_5$, we come close to the most ideal situation.

\vspace{-10pt}
\subsubsection{Effect of tree size}\label{sect.experiment.2}
\vspace{-10pt}

\begin{figure}[t]
  \centering
  
  % Sub-figure (a)
  \includegraphics[width=\linewidth]{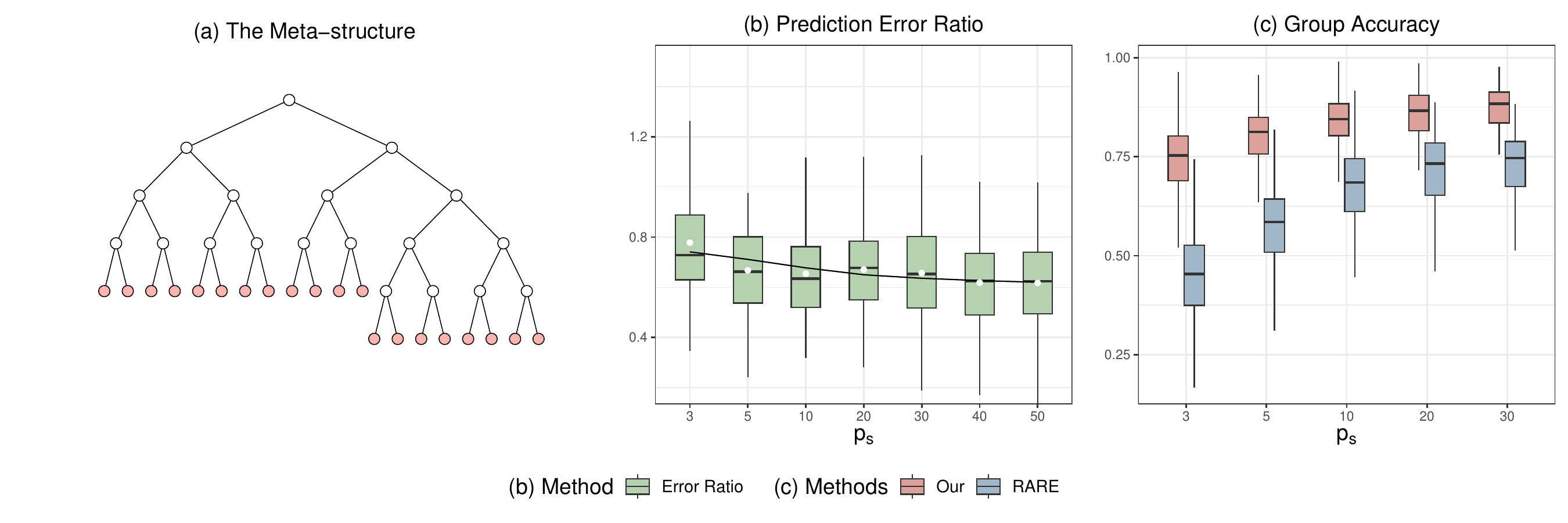}

  \caption{Results of simulation studies described in Section \ref{sect.experiment.2}. (a) The meta-structure from which we generate trees by adding nodes below the colored ones. (b) The boxplot of the ratio between our and RARE prediction errors under 200 replications, the curve fits the average ratio versus $1/\sqrt{\log(p)}$. (c) The contrast of group accuracy of our method and RARE under different tree structures and dimensions.}
  \label{fig.exp2.three_panels}
\end{figure}

Next, we simulate data under the model described in Example 2, Section \ref{exp:fixed-tree}. The number of total groups $K$ is assumed to be fixed, and we evaluate the performance of the methods when the dimension $p$ increases. Figure \ref{fig.exp2.three_panels}a shows the metastructure of our generated trees, with the leaves representing the aggregating set of size $K=20$. Each node in the aggregating set is the root of a replication for a subtree with $p_s$ dimension of features. Therefore, the total dimension $p=20p_s$ and the growing structure below the aggregating set are ignored in the plot. In the simulation, we choose $p_s\in\{3,5,10,20,30,40,50\}$, respectively, so that $p\in\{60,100,200,400,500,800,1000\}$. We train the model on a training set with size 50, and choose the penalty parameter using the square loss on a validation set with the same sample size. The prediction error for our method and the rare feature selection method is evaluated on an independent test set with size 500.

On the theory side, Corollary \ref{corol.tree2} suggests that we should have a finite sample rate $\{\log(p)\}^{-1/2}$ times faster than the rare feature selection method. Though it is merely a description of the error bound, which we are actually unsure about how tight it could be, we can still observe the effect empirically. Figure \ref{fig.exp2.three_panels}b depicts the ratio between prediction errors of our method and the rare method, in 200 replications. We denote the mean value of each setting using the white dots and fit a linear function versus $\{\log(p)\}^{-1/2}$. The black line plots the predicted curve $r=1.0659/\sqrt{\log(p)}+ 0.2145$, with the coefficient significant under the level 0.05. The decreasing trend in the ratio is evident and aligns with the theory. It can also be observed that the ratio is mostly below 1, indicating a lower prediction error than the competitor. In addition, we have the extra bonus of better group selection accuracy, as shown in Figure \ref{fig.exp2.three_panels}c.

\vspace{-10pt}
\subsection{Performance in general settings}\label{sect:general_settings}
\vspace{-10pt}

We further compare our method with the rare feature selection method, under varying configurations. Throughout this section, we simulate the tree with $K$ groups and $p$ features following the simulation procedure in \cite{yan2021rare} using hierarchical clustering. The first $\frac{K}{2}$ groups each has $\frac{3p}{2K}$ features, and the remaining $\frac{K}{2}$ groups each has $\frac{p}{2K}$ features. The effect values of the $K$ distinct groups in $\beta^*$ are randomly obtained from uniform (1.5,2.5) and alternating signs. We then simulate the entries of the design matrix $X\in\mathbb R^{n\times p}$ in i.i.d. Poisson(0.02). Then we consider the following scenarios:
\begin{itemize}[leftmargin=*]
    \item \textbf{Scenario 1}: The outcome $y$ is continuous. The true group number $K$ increases, under a fixed dimension $p$ and sample size $n$. We set $n=50$, $p=100$, and $K\in\{10,20,30,40,50\}$. $y$ is simulated independently in $N(X\beta^*,\sigma^2)$, where $\sigma^2=\|X\beta^*\|_2^2/(5n)$.\vspace{-10pt}
    \item \textbf{Scenario 2}: The outcome $y$ is continuous. The dimension $p$ and group number $K$ increase together at a given ratio $K/p=0.25$. We set $n=500$, let the dimension $p\in\{400,600,800,1000\}$ and $K\in\{100,150,200,250\}$ correspondingly.\vspace{-10pt}
    \item \textbf{Scenario 3}: The outcome $y$ is binary. We set $n=50$, $K=20$, and allow dimension to grow with $p=\{50,100,200,400,600,800,1000\}$. $y$ is generated independently from $y_i\sim$Bernoulli$(1/(1+\exp\{-X_i^T\beta^*\}))$.\vspace{-10pt}
\end{itemize}

\begin{figure}[t]
  \centering
  
  % Row 1
  \includegraphics[width=0.9\linewidth]{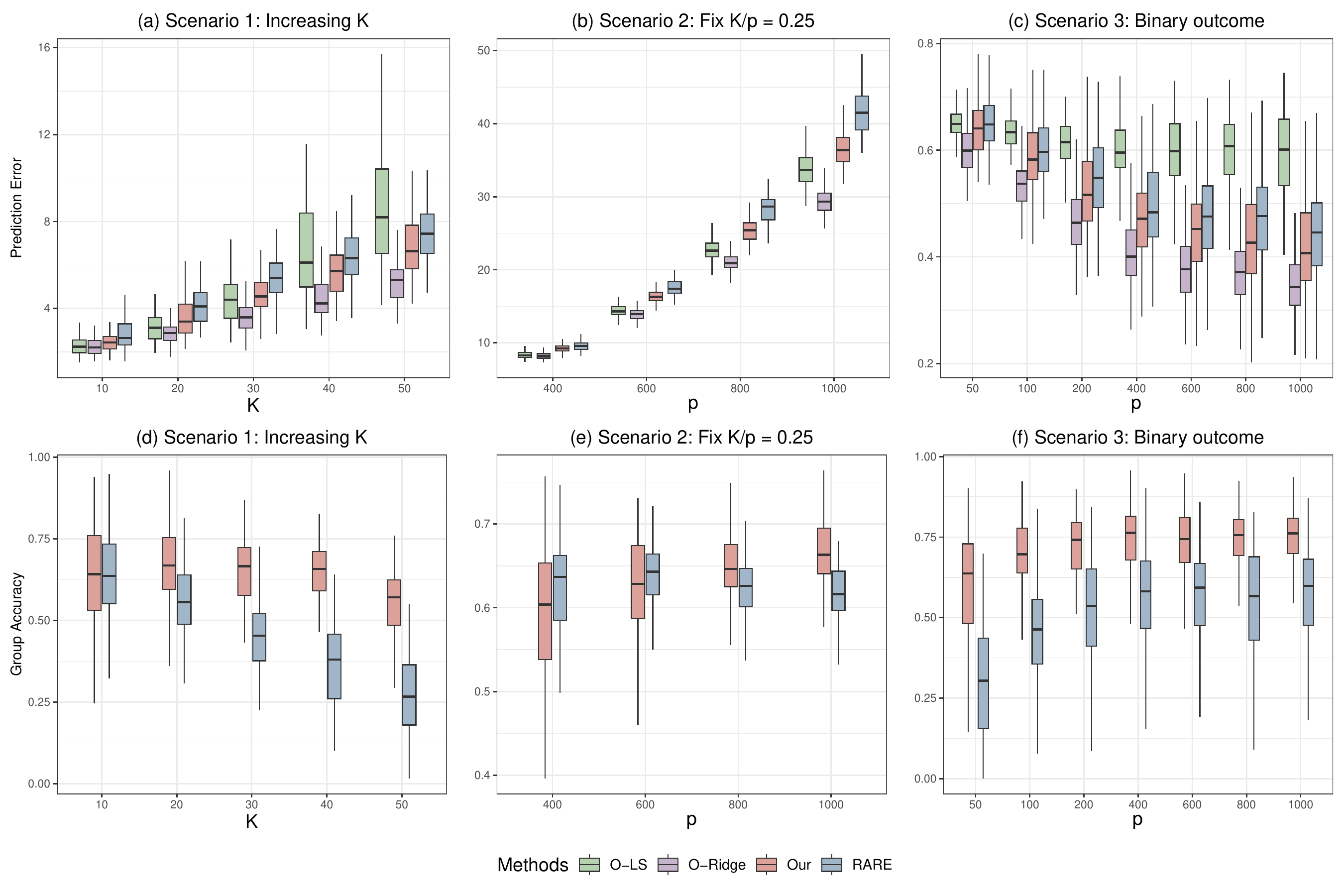}

  \caption{Simulation study results under the three scenarios described in Section \ref{sect:general_settings}. The first row gives the prediction error, and the second row gives the group selection accuracy measured by adjusted Rand index.}
  \label{fig:two-by-three}
\end{figure}

In each scenario, we select the penalty parameter on a validation set with the same sample size $n$ as the training set. The prediction error is then evaluated on an independent test set with size $10n$. The procedure is repeated 200 times, and Figure \ref{fig:two-by-three} summarizes the performances. The results indicate a general advantage in prediction, especially under high-dimensional settings. Scenario 3 further validates the proposed method for classification.

Our advantage in group selection consistency, although not shown theoretically, can be heuristically explained. The rare feature selection method, as a LASSO-type method, does variable selection on the synthetic predictors created by summations over the real predictors under each latent node.  The status of each node is decided individually from the corresponding synthetic predictor. Therefore, for a specific group to be precisely aggregated, all the ancestors of the group latent node have to be simultaneously shrunk to zero, which is generally hard considering the number of nodes. In contrast, our method only requires that the group standard deviation of the corresponding latent node be shrunk to zero. The differences are reflected by a much simpler and smoother solution path (Section \ref{appendix.solution.path}, Supplementary Materials).

In addition, LASSO is well known to be largely affected by the magnitude of predictors in the design matrix, but RARE is not able to standardize before the variable selection step (otherwise, the aggregation is improper). Thus, RARE tends to over-penalize the predictors with smaller magnitude. Therein arises an unfair selection problem between synthetic predictors that are made up of more siblings and fewer siblings within the tree, and it gets more severe when the tree is highly imbalanced. This property also contributes to the less satisfactory grouping accuracy of RARE and is illustrated in the real data analyses.

\vspace{-20pt}
\section{Microbiome data analysis} \label{sect.realdata}
\vspace{-10pt}
We apply the proposed method to analyze microbiome data. As discussed in introduction, we aim to simultaneously choose the aggregation level of OTUs and estimate their group effects specific to each biomedical trait. In this section, we provide a procedure for learning and testing group effects, with examples demonstrating our advantage in scientific discovery.

\vspace{-10pt}
\subsection{Prediction accuracy under relative-shift model}
\vspace{-10pt}

We consider the dataset from \cite{sinha2016fecal}, where researchers collected fecal samples from a total of 131 individuals with colorectal cancer (CRC) and healthy controls. They profiled the microbial communities of the samples (through 16S rRNA gene sequencing) up to the genus level and summarized with a compositional matrix $X\in\mathbb R^{131\times 85}$. A taxonomic tree, as shown in Figure \ref{fig.exp6}, describes the biological relations between the 85 genes. We attempt to predict the continuous outcome, the body mass index (BMI) of each individual; and the binary outcome, the positivity of CRC, respectively. We adopt the relative shift model proposed by \cite{li2023s}, which works directly with the compositional predictors and omits the intercept, i.e., the model assuming 
    ${\rm E}[y_i|X_i]=f^{-1}(\beta_1 X_{i1}+\cdots+\beta_p X_{ip}).$
Compared to the conventional log-contrast model \citep{aitchison1984log}, the relative-shift model directly accommodates zero-valued predictors, and enjoys the property of scale and shift invariance. Each contrast between distinct coefficients $\beta_i-\beta_j$ is interpreted as the relative shift effect between two groups. The model then allows meaningful aggregation of the predictors \citep{li2023s}, it therefore fits perfectly with our framework. We use the squared loss for modeling the BMI, and cross-entropy loss for the CRC outcome.

In each case, we randomly preserve 25\% samples as test set, and select the model on the training set using 5-fold cross-validation. The procedure is repeated 200 times, we compare our method with RARE, plain LASSO and Ridge regression. Table \ref{table.real} shows the prediction loss for CRC and BMI. With continuous outcome, a dramatic improvement in the prediction error can be observed using the tree-guided aggregation methods compared to the usual regularization methods, which provides solid evidence that predictor aggregation is potentially beneficial for prediction tasks. We have a slight improvement compared to the RARE, with average square loss 18.22 versus 18.67. For the binary outcome CRC, there is not much difference between the prediction error or the AUC of the four methods.

%\begin{figure}[H]
%  \centering
  % Sub-figure (a)
    %\includegraphics[width=\linewidth,height=7cm]{fig/experiment5.pdf}
    %\caption{Prediction Loss for BMI and CRC data, witin 200 resamples.}
    %\label{fig.exp5}
%\end{figure}

\begin{table}[t]
\vspace{1.5mm}
  \resizebox{\textwidth}{!}{
  \begin{tabular}{cccccccc}
    \toprule
    \multicolumn{8}{c}{\textbf{Continuous Outcomes}} \\
    \midrule
    \textbf{Data Source} & \textbf{Outcome} & \textbf{Data Size} & \textbf{Est. S--N Ratio} & 
    \textbf{Our Loss} & \textbf{Rare Loss} & \textbf{Lasso Loss} & \textbf{Ridge Loss} \\
    \midrule
    \cite{sinha2016fecal} & BMI & $131\times 85$ & 0.281 & 18.216 & 18.669 & 37.813 & 118.679 \\
     \cite{erawijantari2020influence} & BMI & $96\times1921$ & 0.627 & 9.130 & 9.312 & 549.728 & 160.635 \\
     \cite{wang2020aberrant} & BMI & $287\times2853$ & 0.415 & 16.034 & 16.027 & 46.598 & 531.252 \\
    \cite{yachida2019metagenomic} & BMI & $347\times3418$ & 0.274 & 10.509 & 10.391 & 2328.520 & 209.242 \\
    \midrule
    \multicolumn{8}{c}{\textbf{Binary Outcomes}} \\
    \midrule
    \textbf{Data Source} & \textbf{Outcome} & \textbf{Data Size} & \textbf{Our AUC/Rare AUC} & \textbf{Our Loss} & \textbf{Rare Loss } & \textbf{Lasso Loss } & \textbf{Ridge Loss}  \\
    \midrule
    \cite{sinha2016fecal} & CRC & $131\times 85$ & 0.566/0.576 & 0.637 & 0.632 & 0.625 & 0.635 \\
    \cite{kim2020fecal} & CRC & $240\times121$ & 0.514/0.522 & 0.690 & 0.692 & 0.698 & 0.690 \\
    \cite{jacobs2016disease} & IBD & $90\times507$ & 0.653/0.582 & 0.667 & 0.704 & 0.622 & 0.755 \\
    \cite{erawijantari2020influence} & GC* & $96\times1921$ & 0.551/0.539 & 0.704 & 0.698 & 0.726 & 0.730 \\
    \cite{franzosa2019gut} & IBD*
    & $220\times2834$ & 0.801/0.822 & 0.472 & 0.453 & 0.474 & 0.553 \\
    \cite{wang2020aberrant} & ESRD* & $287\times2853$ & 0.834/0.832 & 0.423 & 0.419 & 0.376 & 0.687 \\
    \cite{yachida2019metagenomic} & CRC & $347\times3418$ & 0.548/0.450 & 0.662 & 0.661 & 0.686 & 0.671 \\
    
    \bottomrule
  \end{tabular}}
  \caption{Prediction results on multiple microbiome datasets with either continuous or binary outcomes.}\label{table.real}
\end{table}

The patterns we discover here turn out to hold true for many other datasets with even higher dimensions, as summarized in Table \ref{table.real}. The table lists the sources of the datasets, along with their outcome types, dimensions, estimated signal-to-noise ratios, and the prediction error means among 200 resamples. The patterns align with the findings of \cite{bien2021tree}, where there is generally not much improvement in prediction using the RARE for microbiome datasets. As was mentioned in both \cite{bien2021tree} and \cite{bichat2020incorporating}, whether it is useful to incorporate taxonomic information into statistical modeling is still an ongoing debate. In addition, a discussion of the impact of signal-to-noise ratio is provided in Supplementary Material \ref{appendix.signal-noise}.%Our method and the analysis for signal strength also serve to provide a new perspective on this problem.

\vspace{-20pt}
\subsection{Post-selection inference on aggregated effects}
\vspace{-10pt}

Another purpose of feature aggregation, to obtain biologically interpretable predictors, often follows by a subsequent analysis on their statistical significance. We take the results on CRC for an example, construct the confidence interval for the aggregated effects, and further illustrate our advantage in dealing with various magnitudes of the predictors.

To avoid double-dipping and conduct post-selection inference, it is common to split the original dataset $S=\{(X_1,y_1),\cdots,(X_n,y_n)\}$ into two parts for model selection and inference, respectively. With the binary outcome CRC, we apply the data fission method proposed by \cite{neufeld2024discussion}. Instead of splitting data points into two independent parts, \cite{leiner2023data} proposed data fission that aims to partition the information and create two datasets each of the original sample size. Specifically, the set $S$ undergoes fission to form two datasets $S_1=\{(X_1,y_1^{(1)}),\cdots,(X_n,y_n^{(1)})\}$ and $S_2=\{(X_1,y_1^{(2)}),\cdots,(X_n,y_n^{(2)})\}$, both with the same design matrix, but have outcomes randomly perturbed in different ways. After fission, the two data sets should together be able to exactly recover the original data set so that no information is lost. In the sparse design context, data fission enjoys a unique advantage of stability by preserving the entire design matrix in both the training and inference sets. \cite{neufeld2024discussion} improves the methodology under the logistic regression setting and allows inference on the exact parameter target, with detailed implementation in Supplementary Material Section \ref{appendix.fission}.

\begin{figure}[th]
  \centering
  % Sub-figure (a)
    \includegraphics[width=0.8\linewidth]{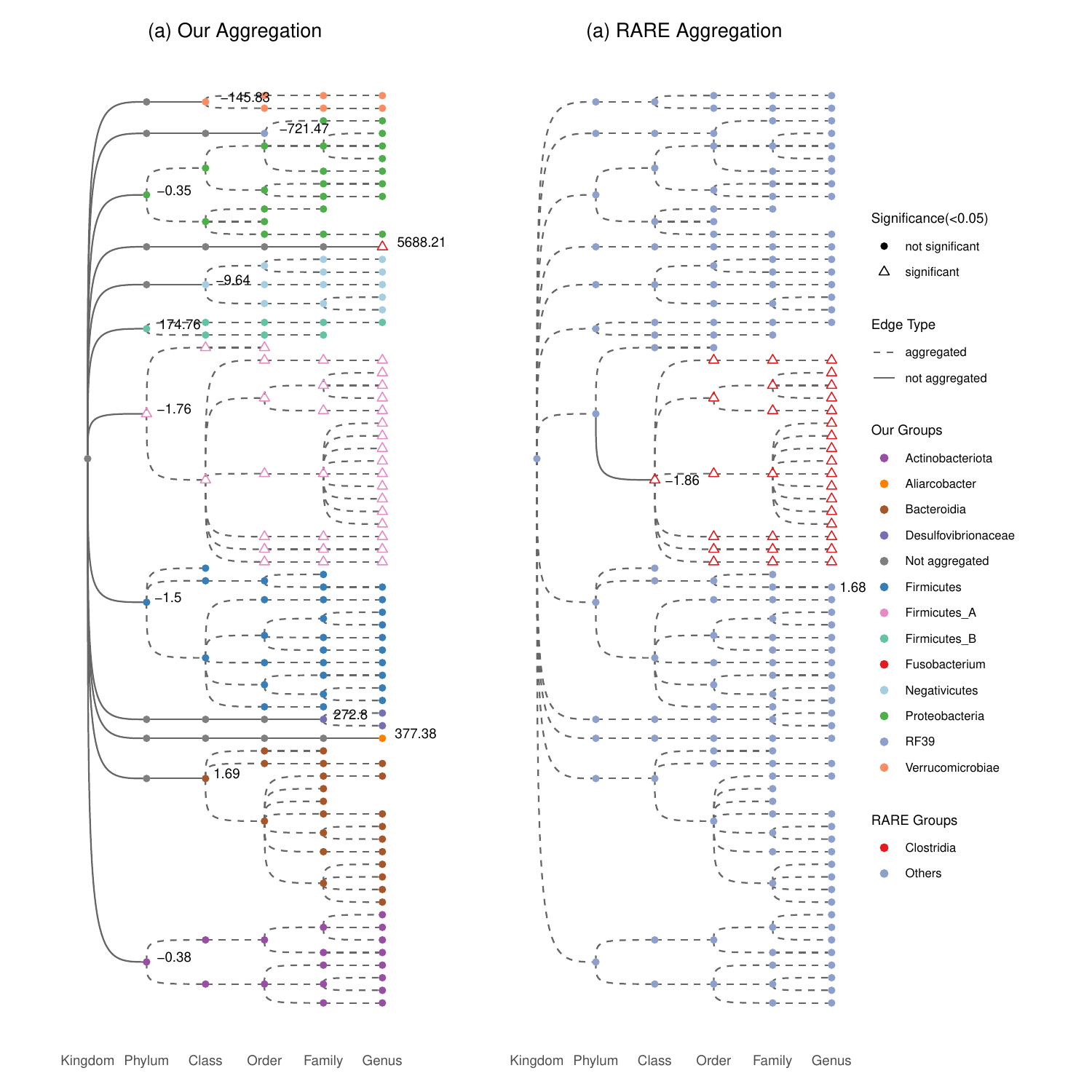}
    \caption{Visualization of the fitted models and results of the post-selection inference procedure based on (a) using our proposed method and (b) RARE \citep{yan2021rare}.}
    \label{fig.exp6}
\end{figure}

 After obtaining the fissured sets, we perform model selection on $S_1$ using 5-fold cross-validation for LASSO, RARE, and our method. Figure \ref{fig.exp6} demonstrates the aggregated predictor groups using different colors. RARE aggregates the predictors into two groups, much fewer than the twelve groups detected by our method. Although the quantities of groups might not have a privilege over one another, the problem is that RARE, not being a group-wise penalty, violates the tree structure to obtain the highly aggregated groups. 

We then select or aggregate the predictors in $S_2$ based on the models selected in the first step, and fit the standard logistic regression using the offsets provided as in \eqref{mid.log-odds}. Inference is carried out on the fitted model for those aggregated or selected predictors, whose fitted values and significance are also labeled in Figure \ref{fig.exp6}.

\vspace{-20pt}

While both RARE and our method detect one significant taxonomic group Clostridia, we have another significant feature, Fusobacterium at Genus level, that is overlooked by RARE. The OTU Fusobacterium is not only identified by the LASSO procedure to be statistically significant, but also a well-studied type of bacteria that is highly associated with CRC in biomedical studies \citep{kostic2013fusobacterium}. We also conduct a multiple testing on the effect contrasts between Fusobacterium and the other eleven aggregated gorups. The p-values, after Benjamini-Hochberg correction procedure\citep{benjamini1995controlling}, are shown in Table \ref{table.contrast.p} to be all significant. The OTU Fusobacterium is noticably rare within the sample, only 17 out of 131 individuals have a nonzero value on this feature. Meanwhile, the tree is highly imbalanced, not allowing it to be aggregated with any other predictors. As a result, it participates in variable selection with an overly small magnitude in the design matrix, and is therefore screened out by RARE to be aggregated with another group. \medskip

\vspace{-10pt}
\begin{table}[H]
\centering
\begin{tabular}{*{5}{c}}
\toprule
$\textbf{Groups}$ & Bacteroidia & Firmicutes & Proteobacteria & Actinobacteriota \\ 
\midrule
% ---- first data row ----
\textbf{p-value} & 0.0090 & 0.0090 & 0.0090 & 0.0090  \\
\bottomrule
$\textbf{Groups}$ & Aliarcobacter & Negativicutes & Desulfovibrionaceae & Firmicutes\_B \\
\midrule
\textbf{p-value} & 0.0128 &
0.0090 & 0.0128 & 0.0100\\
\bottomrule
$\textbf{Groups}$& Verrucomicrobiae & RF39 & Firmicutes\_A & \\
\midrule
\textbf{p-value} & 0.0090 & 0.0128 & 0.0090 &\\
\bottomrule
\end{tabular}
\caption{The adjusted p-values of the group effect contrasts with Fusobacterium after B-H correction, inferred on $S_2$.}
\label{table.contrast.p}
\end{table}
%\vspace{-30pt}
%\subsubsection*{Acknowledgements}
%\vspace{-10pt}
%J. Fu and H. Zou's contributions were supported by a grant from the National Science Foundation (CNS 2220286). A. J. Molstad's contributions were also supported by a grant from the. National Science Foundation (DMS 2413294). 

\vspace{-20pt}
\subsubsection*{Data Availability Statement}
\vspace{-10pt}
The authors confirm that the data supporting the findings of this study are available in its supplementary materials, and at \url{https://github.com/JinwenFu001/direct_feature_aggregation}.

\vspace{-10pt}

\setstretch{1}
%\section*{Acknowledgments}
%A. J. Molstad's research was supported in part by a grant from the National Science Foundation (DMS-2415067). 
%The authors used ChatGPT (versions o1 Pro and o3-Mini-High) to assist with proofreading, code development, and code documentation. All content and conclusions in this manuscript remain the sole responsibility of the authors. The authors verified and approved all AI-assisted suggestions prior to including them in the final manuscript.
\spacingset{1} 
% \section*{Supplementary material}

% This supplementary file contains proofs of all theoretical results, detailed derivations and pseudocode for the accelerated proximal-gradient estimator, tuning-parameter guidelines, auxiliary lemmas, and extended simulation results.

\singlespacing
\vspace{-10pt}
\bibliographystyle{abbrvnat}
\singlespacing
\bibliography{reference}

\end{document}